\documentclass[10pt,journal]{IEEEtran}
\usepackage[caption=false,font=normalsize,labelfont=sf,textfont=sf]{subfig}
\usepackage{enumitem}
\usepackage{graphicx}
\usepackage{tabularx}
\usepackage{multirow}
\usepackage{csquotes}
\usepackage{url}
\usepackage{booktabs}
\usepackage{hyperref}
\usepackage{listings}
\usepackage{adjustbox}
\usepackage{amsmath}
\usepackage{changepage}
\usepackage{listings}
\usepackage{calc}
\usepackage{multirow}
\usepackage{pdflscape}
\usepackage{tcolorbox}
\usepackage{breqn}
\usepackage{orcidlink}

\newcommand{\tbl}[1] { \textcolor{black}{#1}}

\usepackage{balance}
\usepackage{makecell} % for more vertical space in cells
\setcellgapes{5pt}

\title{RLocator: Reinforcement Learning for Bug Localization}

\author{
    \IEEEauthorblockN{Partha Chakraborty\orcidlink{0000-0001-5965-615X}, \IEEEmembership{Student Member, IEEE}, Mahmoud Alfadel\orcidlink{0000-0002-2621-6104}, \IEEEmembership{Member, IEEE},  and Meiyappan Nagappan\orcidlink{0000-0003-4533-4728}~\thanks{Partha Chakraborty and Meiyappan Nagappan is with David R. Cheriton School
of Computer Science, University of Waterloo,  Waterloo, Canada, N2L 3G1. \\E-mail:\{p9chakra, mei.nagappan\}@uwaterloo.ca}
~\thanks{
Mahmoud Alfadel is with the Department of
of Computer Science, University of Calgary,
Calgary, Canada, T2N 1N4. \\E-mail:\ mahmoud.alfadel@ucalgary.ca}
}
}

\begin{document}

\IEEEtitleabstractindextext{
\begin{abstract}
   Software developers spend a significant portion of time fixing bugs in their projects. To streamline this process, bug localization approaches have been proposed to identify the source code files that are likely responsible for a particular bug. 
Prior work proposed several similarity-based machine-learning techniques for bug localization. 
Despite significant advances in these techniques, they do not directly optimize the evaluation measures.
We argue that directly optimizing evaluation measures can positively contribute to the performance of bug localization approaches.
% Prior studies  optimize similarity metrics in the training phase, while in the testing phase, they evaluate using ranking metrics.

Therefore, in this paper, we utilize Reinforcement Learning (RL) techniques to directly optimize the ranking metrics.
% leading to an efficient  bug localization model.
% In this paper, 
We propose \textsc{RLocator}, a Reinforcement Learning-based  bug localization approach. 
We formulate RLocator using a Markov Decision  Process (MDP) to optimize the evaluation measures directly. 
%In each time step of MDP, RLocator ranks the source code files based on their possibility of being responsible for a bug.
We present the technique and experimentally evaluate it based on a benchmark dataset of 8,316 bug reports from six highly popular Apache projects. 
The results of our evaluation reveal that RLocator achieves a Mean Reciprocal Rank (MRR) of 0.62, a Mean Average Precision (MAP) of 0.59, and a Top 1 score of 0.46.
We compare RLocator with three state-of-the-art bug localization tools, FLIM, BugLocator, and BL-GAN. 
Our evaluation reveals that RLocator outperforms both approaches by a substantial margin, with improvements of 38.3\% in MAP, 36.73\% in MRR, and 23.68\% in the Top K metric.
These findings highlight that directly optimizing evaluation measures considerably contributes to performance improvement of the bug localization problem.

% Software developers spend a significant portion of their work time fixing bugs. Bug localization tools can speed up the process by shortlisting potentially responsible source code files for a particular bug.
% Prior work proposed several similarity-based supervised machine-learning techniques for bug localization. 
% Despite significant recent advances in these techniques, they do not directly optimize the evaluation measures. Instead, they use different metrics in the training and testing phases, which can negatively impact the model performance in
% retrieval tasks.
%
% In this paper, we propose \emph{RLocator}, a Reinforcement Learning-based (RL) bug localization approach. 
% We formulate the bug localization problem using a Markov Decision  Process (MDP) to optimize the evaluation measures directly. 
% In each time step of MDP, RLocator ranks the source code files based on their possibility of being responsible for a bug. 
% We formally present the technique and experimentally evaluate it based on a benchmark dataset of more than 9,000 bug reports from six highly popular Apache projects.
% Our evaluation shows that RLocator achieves a high Mean Reciprocal Rank (MRR) of $0.61$ and Mean Average Precision (MAP) of $0.59$, outperforming seven state-of-the-art bug localization techniques.
% Such results demonstrate that direct optimization of evaluation measures considerably contributes to performance improvement of the bug localization problem.

\end{abstract}
}
\IEEEpubid{\copyright~2024 IEEE. Author pre-print copy. The final publication is available online at: \url{https://doi.org/10.1109/TSE.2024.3452595}}
\maketitle
\IEEEdisplaynontitleabstractindextext
\section{Introduction}
\label{sec:intro}
Software bugs are an inevitable part of software development. 
Developers spend one-third of their time debugging 
 and fixing bugs~\cite{LaToza2010}.
After a bug report/issue has been filed, the project team identifies the source code files that need to be inspected and modified to address the issue. 
However, manually locating the files responsible for a bug is expensive (in terms of time and resources), especially when there are a lot of files and bug reports. 
Moreover, the number of bugs reported is often higher than the number of available developers~\cite{Anvik2005}. Consequently, the fix-time and maintenance costs rise when the customer satisfaction rate decreases~\cite{Zhou2012}.

% Prior studies have proposed various information retrieval-based bug localization (IRBL)

Bug Localization is a method that refers to identifying the source code files where a particular bug originated.
% Bug localization techniques  can shortlist the source code files potentially responsible for a bug based on the report's description.
Given a bug report, bug localization approaches utilize the textual information in the bug report, and the project source code files to shortlist the potentially buggy files.
Prior work has proposed various Information Retrieval-based Bug Localization (IRBL) approaches to help developers speed up the debugging process (e.g., Deeplocator~\cite{Xiao2018}, CAST~\cite{Liang2019}, KGBugLocator~\cite{Zhang2020}, BL-GAN~\cite{Zhu2022}).
%
% ~\cite{Li2019, Xiao2018, Liang2019, Lam2015, Zhu2021, peter2018, Zhang2020}. 
% A bug localization tool can shortlist the source code files potentially responsible for a bug based on the report's description.
% Given a bug report, BL approaches utilize the textual information in the bug report and the source code files to shortlist the potentially buggy source code files. 
% A plethora of prior work focused on proposing solutions to improve the performance of IRBL. 
% An example of Deeplocator~\cite{Xiao2018} and CAST~\cite{Liang2019} use AST of the source code, Pathidea~\cite{Chen2022} utilizes logs from the bug report and DEMOB~\cite{Zhu2021} applies an attention mechanism to localize bugs.
% \par \noindent.

One common theme among these approaches is that they follow a similarity-based approach to localize bugs. Such techniques measure the similarity between bug reports and the source code files. For estimating similarity, they use various methods such as cosine distance~\cite{Ciborowska2022}, Deep Neural Networks (DNN)~\cite{Lam2015}, and Convolutional Neural Networks (CNN)~\cite{Liang2019}. 
Then, they rank the source code files based on their similarity score.
In the training phase of these approaches, the model learns to optimize the similarity metrics.
In contrast, in the testing phase, the model is tested with ranking metrics (e.g., Mean Reciprocal Rank (MRR) or Mean Average Precision (MAP)).

While most of these approaches showed promising performance, they optimize a metric that indirectly represents the performance metrics. Prior studies~\cite{Wei2017, Alejo2010, Xu2017, Yue2007} found that direct optimization of evaluation measures substantially contributes to performance improvement of ranking problems. Direct  optimization is also efficient compared to optimizing indirect metrics~\cite{Yue2007}. 
Hence, we argue that it is challenging for the solutions proposed by prior studies to  sense how a wrong prediction would affect the performance evaluation measures~\cite{Wei2017}.
% using evaluation measures in 
% %We argue that using evaluation measures  in 
% optimization will also improve the performance of bug localization tools.
In other words, if we use the retrieval metrics (e.g., MAP) in the training phase, the model will learn how each prediction will impact the evaluation metrics. 
A wrong prediction will change the rank of the source code file and ultimately impact the evaluation metrics.
% Discuss MDP + RL

Reinforcement Learning (RL) is a sub-category of machine learning methods where labeled data is not required. 
In RL, the model is not trained to predict a specific value. Instead, the model is given a signal about a right or wrong choice in training~\cite{Sutton2018}. 
Based on the signal, the model updates its decision.
This allows RL to use evaluation measures such as MRR and MAP in the training phase and directly optimize the evaluation metrics. 
Moreover, because of using MRR/MAP as a signal instead of a label, the problem of overfitting will be less prevalent.
Markov Decision Process (MDP) is a foundational element of RL. 
MDP is a mathematical framework that allows the formalization of discrete-time decision-making problems~\cite{Garcia2013}.
Real-world problems often need to be formalized as MDP to apply RL.
\newpage

In this paper, we present \emph{RLocator}, an RL technique for localizing software bugs in source code files.
We formulate RLocator into an MDP.
In each step of the MDP,  we use MRR and MAP  as signals to guide the model to optimal choice.
%
% - We use this x of projects to evaluate
% - We got this MAP (x - x) MRR (x - x) for these
% - Our results shows that our represent a significant advance over the state-of-the-art by comparing RLocator with x approaches.
% - Our results show that RLocator outperforms SOA approaches, e.g., RLocator outperforms the best competitor by 1%-x% in all the studied projects.
%
% \mei{I would remove the rest of the para and move some of the salient points into contributions. Otherwise it is repetitive.}
% We evaluate RLocator on a set of 8,316 bug reports belonging to six open-source Apache projects.
% RLocator achieves an MRR of 0.49 - 0.61 across all studied projects.
% For MAP, RLocator achieves 0.47 - 0.59.
% In six projects, 0.61 is the highest MRR RLocator achieves. (0.49 - 0.61), and 0.59 is the highest MAP (0.47 - 0.59) achieved by RLocator.
We evaluate RLocator on a benchmark dataset of six Apache projects and find that, compared with existing state-of-the-art bug localization techniques, RLocator achieves substantial performance improvement. 
While pinpointing the exact reasons for RL's superior performance over other supervised techniques can be challenging, RL learns more generalizable approaches, especially in dynamic and complex environments. In comparison to supervised learning, it learns approaches that are more adaptable to a variety of situations~\cite{GDI2021, Kober2013}, which is a form of generalization.  Additionally, RL demonstrates proficiency in scenarios where the optimal solution is not clearly defined, showcasing its versatility across various tasks and domains~\cite{Sutton2018}. These factors can contribute to the superior performance of RLocator.\\

% The evaluation result shows that RLocator is effective. For example, in the AspectJ project, RLocator achieved MRR of $0.59$, which means, on average, the buggy file is ranked on the $1/0.59 \thickapprox 2^{nd}$ position. Our evaluation also shows that
% RLocator outperforms existing bug localization methods BugLocator~\cite{Zhou2012}, DeepLoc~\cite{Xiao2019}, CAST~\cite{Liang2019}, KGBugLocator~\cite{Zhang2020}, BL-GAN~\cite{Zhu2022}, FBL-BERT~\cite{Ciborowska2022}, and FLIM~\cite{Liang2022} in terms of Mean Average Precision.
% \par \noindent

The main contributions of our work are as follows:
\begin{itemize}
    \item We present RLocator, an RL-based software bug localization approach. The key technical novelty of RLocator is using RL for bug localization, which includes formulating the bug localization process into an MDP. 
 
    \item We provide an experimental evaluation of RLocator with 8,316 bug reports from six Apache projects. When RLocator can localize, it achieves an MRR of 0.49 - 0.62, MAP of 0.47 - 0.59, and Top 1 of 0.38 - 0.46 across all studied projects. Additionally, we compare RLocator's performance with state-of-the-art bug localization methods. RLocator outperforms FLIM~\cite{Liang2022} by 38.3\% in MAP, 36.73\% in MRR, and 23.68\% in Top K. Furthermore, RLocator exceeds BugLocator~\cite{Zhou2012} by 56.86\% in MAP, 41.51\% in MRR, and 26.32\% in Top K. In terms of Top K, RLocator shows improved performance over BL-GAN~\cite{Zhu2022}, with gains ranging from 55.26\% to 3.33\%. The performance gains for MAP and MRR are 40.74\% and 32.2\%, respectively.
\end{itemize}

\section{Motivation}
Reinforcement Learning (RL) stands out for its ability to learn from feedback, a characteristic that empowers models to self-correct based on the outcomes of their actions. This feature finds widespread application, exemplified by platforms like Spotify, an audio streaming service using RL to learn user preferences~\cite{Maystre2023}. The model evolves and adapts by presenting music selections and refining recommendations through user interactions. The versatility of RL extends beyond entertainment; various companies~\cite{Chen2018} and domains~\cite{Yu2021} leverage its capacity for iterative learning and adjustment.

The proficiency required for bug localization is often acquired through experience, with seasoned developers exhibiting a faster bug-finding aptitude than their less experienced counterparts~\cite{Winter2023}. Recognizing the significance of experience in bug localization, we propose the integration of reinforcement learning into this domain. By employing RL, a model can present developers with sets of source code files as possible causes for a bug and learn from their feedback to enhance its skill in localizing bugs in the software. In contrast to conventional machine learning approaches, which rely solely on labeled data and lack easy adaptability, reinforcement learning presents two distinct advantages: firstly, the ability to learn from developer feedback, and secondly, the elimination of the requirement for labeled data in real-world scenarios. Therefore, our research aims to incorporate reinforcement learning into bug localization, leveraging its capacity to adapt and enhance performance through iterative feedback.

% To help illustrate how our approach works and the challenges of using typical bug localization systems we provide a simple motivating example.\par
% Amy is a developer who has started working on a new project. Her organization uses a bug localization tool however the tool is useless in Amy's project as the project has just started.\par

% Our proposed technique aims to be effective in scenarios like Amy's. In the bug localization process developers like Amy will go through the source code file and assess whether a particular file is responsible for a bug.

% Rlocator will suggest a set of files at first and Amy can manually verify whether the suggestion was helpful in locating the bugs. From Amy's feedback (relevant or not relevant) Rlocator will learn to suggest more relevant files.
\section{Background}
\label{sec:background}
In this section, we describe terms related to the bug localization problem, which 
we use throughout our study.
% to define the bug localization problem in our work,
Also, we present an overview of reinforcement learning.

\subsection{Bug Localization System}
\label{subsec:bg_bl_system}
% Bug Localization tools speed up the debugging process. The required time to manually localize bugs may depend on the codebase's scale and complexity. A bug in a large and complex codebase will take more time than a bug originated from a small codebase. Moreover, the required time also depends on the developers' expertise~\cite{Sharma2019}. A new developer will undoubtedly take longer than a developer already accustomed to the codebase. The required time is crucial regarding the commercial aspect of software development.\par
A typical bug localization system utilizes several sources of data, e.g., bug reports, stack traces, and logs, to identify the responsible source code files. One particular challenge of the system is that the bug report contains natural language, whereas source code files are written in a programming language.\par
Typically, bug localization systems identify whether a bug report relates to a source code file.  
To do so, the system extracts features from both the bug report and the source code files. 
Previous studies  used techniques such as N-gram~\cite{Wang2016, Miryeganeh2021} and Word2Vec~\cite{Kim2020, Chen2020} to extract features (embedding) from bug reports and source code files. Other studies (e.g., Devlin et al.~\cite{devlin2019}) introduced the transformer-based model BERT which has achieved higher performance than all the previous techniques. One of the reasons transformer-based models perform better in extracting textual features is that the transformer uses multi-head attention, which can utilize long context while generating embedding. Previous studies have proposed a multi-modal BERT model~\cite{feng2020} for programming languages, which can extract features from both bug reports and source code files.
% In our case, the source code files are long. 
% Thus, a method declared far away from its usage may not be captured in Word2Vec. However, because of the long context, BERT can utilize that information. Moreover, Word2Vec generates embedding for a word, which will not be updated in the inference phase based on the word's context. Contrary, the BERT model will generate the embedding for a sequence. Thus, it can utilize the context in the inference phase too. In source code files, the use of a variable depends on the scope. The variable name can be declared again in another scope and may contain different objects. As the BERT model generates embedding for sequence, the embedding will be different for each use of the variable. Previous studies have proposed a multi-modal BERT model~\cite{feng2020} for programming languages, which can extract features from both bug reports and source code files.

%  \begin{figure}[tb!]
%     \centering    \fbox{\includegraphics[scale=0.024]{Figures/Bug_Example.png}}
%     \caption{Sample bug report.}
%     \label{fig:sample report}
% \end{figure}
A bug report mainly contains information related to unexpected behavior and how to reproduce it. It mainly includes a bug ID, title, textual description of the bug, and version of the codebase where the bug exists. The bug report may have an example of code, stack trace, or logs. 
A bug localization system retrieves all the source code files from a source code repository at that particular version. For example, assume we have 100 source code files in a repository in a specific version. After retrieving 100 files from that version, the system will estimate the relevance between the bug report and each of the 100 files. The relevance can be measured in several ways. For example, a naive system can check how many words of the bug report exist in each source code file. A sophisticated system can compare embeddings using cosine distance~\cite{Wang2020sim}. After relevance estimation, the system ranks the files based on their relevance score. The ranked list of files is the final output of a bug localization system that developers will use.
% A deep learning-based system can use a neural network to estimate the relevance between the embeddings of source code files and bug reports. The neural network output will be scaled relevance between the bug report and the source code file. The ranked list of files is the final output of a bug localization system that developers will use.

\subsection{Reinforcement Learning}
In Reinforcement Learning (RL), the agent interacts with the environment through observation. Formally, an observation is called \enquote{State,} $S$. In each state, at time $t$, $S_t$, the agent takes action $A$ based on its understanding of the state. 
Then, the environment provides feedback/reward $\Re$ and transfers the agent into a new state $S_{t+1}$. The agent's strategy to determine the best action, which will eventually lead to the highest cumulative reward, is referred to as \emph{policy}~\cite{Fujimoto2019, Sutton2018}.\par
The cumulative reward (until the goal/end) that an agent can get if it takes a specific action in a certain state is called \emph{Q value}.
The function that is used to estimate the \emph{Q value} is often referred as \emph{Q function} or \emph{Value function}.\par
In RL, an agent starts its journey from a starting state and then goes forward by picking the appropriate action. The journey ends in a pre-defined end state. The journey from start to end state is referred to as \emph{episode}.\par
From a high level, we can divide the state-of-the-art RL algorithms into two classes. The first is the model-free algorithms, where the agent has no prior knowledge about the environment. The agent learns about the environment by interacting with the environment. The other type is the model-based algorithm. In a model-based algorithm, the agent uses the reward prediction from the model instead of interacting with the environment.

The bug localization task is quite similar to the model-free environment as we cannot predict/identify the buggy files without checking the bug report and source code files (without interacting with the environment). Thus, we use model-free RL algorithms in this study. 
Two popular variants of model-free RL algorithms are: 
\begin{itemize}
    \item \emph{Value Optimization}:
The agent tries to learn the Q value function in value optimization approaches. 
The agent keeps the Q value function in memory and updates it gradually.
It consults the Q value function in a particular state and picks the action that will give the highest value (reward). 
An example of the value optimization-based approach is Deep Q Network (DQN)~\cite{Sutton2018}.
\item \emph{Policy Optimization}:
In the Policy optimization approach, the agent tries to learn the mapping between the state and the action that will result in the highest reward. The agent will pick the action based on the mapping in a particular state. 
An example of the policy optimization-based approach is Advantage Actor-Critic (A2C)~\cite{Nguyen2021, Sutton2018}.
\end{itemize}
A2C is a policy-based algorithm where the agent learns an optimized policy to solve a problem.
In Actor-Critic, the actor-model picks action. The future return (reward) of action is estimated using the critic model. The actor model uses the critic model to pick the best action in any state. Advantage actor-critic subtracts a base value from the return in any timestep. 
A2C with entropy adds the entropy of the probability of the possible action with the loss of the actor model. 
As a result, in the gradient descent step, the model tries to maximize the entropy of the learned policy. 
Maximization of entropy ensures that the agent assigns almost an equal probability to an action with a similar return.

% XGBoost (Extreme Gradient Boosting) is a machine learning algorithm that utulizes gradient boosting technique to build efficent treaa based classiifer. XGBoost iteratively build week decision tree based predictive model where each tree segment only focusses on fixing the wrong prediction of the previous  tree.
% % \input{Dataset}
\section{RLocator: Reinforcement Learning for Bug Localization}
\label{sec:ranking}
 \begin{figure*}[ht!]
    \centering   
    \fbox{
        \includegraphics[scale=0.27]{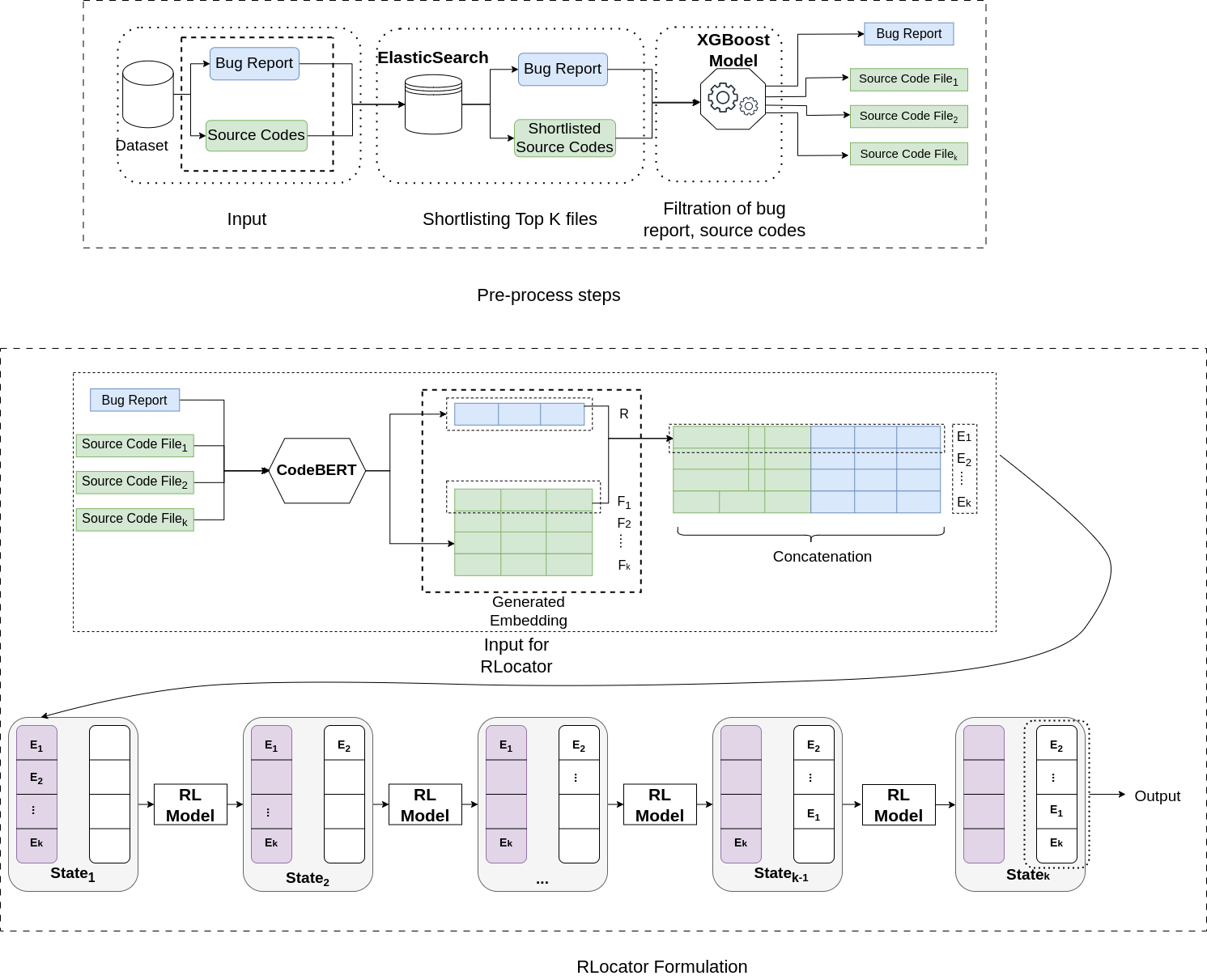}
    }
    \caption{Bug Localization as Markov Decision Process.}
    \label{fig:bl_as_mdp}
\end{figure*}
In this section, we discuss the steps we follow to use RLocator. 
% Our approach is comprised of two main steps. 
% In section~\ref{sec: identification_of_unrankable_data}, 
First, we explain (in Section~\ref{sec: identification_of_unrankable_data}) the pre-processing step required for using RLocator.
% the filtration of data, 
Then, we explain (in Section~\ref{sec:formulation_of_RLocator}) the formulation steps of our design of RLocator. 
We present the overview of our approach in Figure~\ref{fig:bl_as_mdp}. 
\subsection{Pre-process}
Before using the bug reports and source code files to train the RL model, they undergo a series of pre-processing steps. The steps are described in this section.\\
\label{sec: identification_of_unrankable_data}
\textbf{Input:}The inputs to our bug localization tool are bug reports and source code files associated with a particular version of a project repository. 
Software projects maintain a repository for their bugs or issues (e.g., Jira, Github, Bugzilla). 
The first component, the bug report, can be retrieved from those issue repositories. 
We use the bug report to obtain the second component (i.e., source code) by identifying the project version affected by the bug. Typically, each bug report is associated with a version or commit SHA of the project repository. After identifying the buggy version, we collect all source code files from the specific version of the code repository. In the training phase, we compile bug reports and source code files into a dataset for subsequent usage. Our dataset contains a set of bug reports where each bug report has its own set of source code files. In real-world usage, RLocator directly accesses bug reports and source code files from the repository. In Figure~\ref{fig:bl_as_mdp}, we illustrate the input stage where we get the bug report and source code files from the dataset.

\noindent\textbf{Shortlisting source code files:} The number of source code files in different versions of the repository can be different. All of the source code files can be potentially responsible for a bug. In this step, we identify $K$ source code files as candidates for each bug. We limited the candidates to $K$ as we cannot pass a variable number of source code files to the RL model. Moreover, given that RLocator primarily learns from developers' feedback, its usage can prove challenging for a developer with many candidate source code files. To illustrate the issue, consider a repository with 700 files. RLocator presents files to the developer one by one for relevance verification. This sequential approach significantly prolongs the time taken to find a relevant file, resulting in a waste of developers' time. Consequently, it is crucial to limit the number of files shown to developers by providing a shortlisted set for assessment.\\
To identify the $K$ most relevant files, we use ElasticSearch (ES).  
ES is a search engine based on the Lucene search engine project. It is a distributed, open-source search and analytic engine for all data types, including text. It analyzes and indexes words/tokens for the textual match and uses BM25 to rank the files matching the query. We use the ES index for identifying the topmost k source code files related to a bug report. Following the study by Liu et el.~\cite{Liu2022}~(who used ES in the context of code search), we build an ES index using the source code files and then queried the index using the bug report as the query.
Then, we picked the first $k$ files with the highest textual similarities with the bug report. 
We want to note that the goal of bug localization is to get the relevant files to be ranked as close to the $1^{st}$ rank as possible. Hence, metrics like MAP and MRR can measure the performance of bug localization techniques. 
While one can argue why we not only rely on ES to rank the relevant files, we find that the MAP and MRR of using ES are poor. Our RL-based technique learns from feedback and aims to rerank the output from ES to get higher MAP and MRR scores. In Figure~\ref{fig:bl_as_mdp}, we illustrated the candidate refinement step where we query ElasticSearch using the bug report and use outputs to refine the candidate source code files.

\noindent \textbf{Filtration of bug report and source code files:} One limitation of ES is that it sometimes returns irrelevant files among the top $k$ most relevant source code files. When there are no relevant files in the first $k$ files, it hinders RLocator training using developer feedback and introduces noise. Therefore, we use an XGBoost-based binary classifier~\cite{Chen2016} to identify cases where ES may return no relevant files in the top $k$ files. The rationale for using XGBoost is twofold: (1) to optimize developer time by not presenting irrelevant files and (2) to filter out noise during training.

ES-based filtering is not used because its similarity values are not normalized, and cosine similarity is inapplicable to text data. We provide the XGBoost model with the bug report and the top $k$ files retrieved by ES to determine if any are relevant. If the XGBoost model predicts no relevant files in the set, we exclude those bug reports and their associated files. Each bug report is associated with its unique set of source code files, so filtering one does not impact others.

% One limitation of ES is that, in some cases, it returns irrelevant files in the $k$ most relevant source code files. 
% As there are no relevant files in the first $k$ files, we cannot use those files to train RLocator using developers' feedback. Also, including such data in the model induces noise in the training process. Therefore, we employ an XGBoost-based binary classifier model~\cite{Chen2016} to identify the cases where ES may return no relevant files in the top $k$ files. In short, the rationale for using XGBoost is two fold: (1) to optimize developer time (by not presenting the set of files where there are no relevant files) and (2) to filter out as much noise as possible in the training process. We do not use ES-based filtering because its similarity values are not normalized. Additionally, cosine similarity is not applicable here since it only works with vector data, whereas we have text data (source code and bug reports). We provide the model with the bug report and the top $k$ relevant files retrieved by ElasticSearch (ES). The model's objective is to determine whether the ES output, comprising a set of files for each bug report, includes any pertinent source code files. Using the prediction from the XGBoost model, we exclude bug reports (along with their corresponding source code files) in cases where the ES output lacks relevant files. It is important to highlight that each bug report is associated with its unique set of source code files; the filtration of a bug report and its associated files does not impact other bug reports.
\setlength{\tabcolsep}{3pt}
\begin{table}[tb!]
\centering
\caption{Description and rationale of the selected features.}
\begin{tabular}{@{}lp{3cm}p{3cm}@{}}
\toprule
\textbf{Feature}   & \textbf{Description}                                                                                                                                                     & \textbf{Rationale}                                                                                                                                                                                                                                              \\ \midrule
Bug Report Length  & Length of the bug report.                                                                                                                                                & Fan et al.~\cite{Fang2021} found that it is hard to localize bugs using a short bug report. A short bug report will contain little information about the bug. Thus it will be hard for the ElasticSearch to retrieve source code file responsible for this bug. \\ \midrule
Source Code Length & Median length of the source code files associated with a particular bug. Note that we calculate the string length of the source code files after removing code comments. & Prior studies~\cite{Lv2011, Lewis1998} found that calculating textual similarity is challenging for long texts. Length of source code may contribute to the performance drop of ElasticSearch.                                                                  \\ \midrule
Stacktrace         & Availability of stack trace in bug report.                                                                                                                               & Schroter et al.~\cite{Schroter2010} found that stacktraces in bug reports can help the debugging process as they may contain useful information. Availability of stacktraces may improve the performance of ElasticSearch.                                      \\ \midrule
Similarity         & Ratio of similar tokens between a bug report and source code files                                                                                                       & Similarity indicates the amount of helpful information in the bug report. We calculate the similarity based on the equation presented in Section~\ref{sec:results}.                                                                                             \\ \bottomrule
\end{tabular}
\label{table:feature_description}
\end{table}
To build the model, we study the most important features associated with the prediction task.
We consult the related literature on the field of information retrieval~\cite{Lv2011, Lv2011_plus, Lewis1998} and bug report classification~\cite{Fang2021} for feature selection. 
The list of computed features is presented in Table~\ref{table:feature_description}.
For our dataset, we calculate the selected features and trained the model using 10-fold cross-validation.
The results show that our classifier model has a precision of 0.78, a recall of 0.93, and an F1-score of 0.85. 
Additionally, the model is able to correctly classify 91\% of the dataset (there will be relevant source code files in the top $k$ files returned by ES). 

%In the experimental setting, we first query ES. After receiving the top $k$ files from ES, we pass the bug report and the $k$ source code files to the XGBoost model. XGBoost model predicts whether there are any relevant files in the top $k$ files or not. 

After filtration, we pass each bug report and its source code files to RLocator.
In Figure~\ref{fig:bl_as_mdp}, we have depicted the operational procedure of RLocator. The workflow commences with a curated dataset containing bug reports and source code files. Subsequently, we index the source code files into the ES index. From ES, we obtain bug reports and shortlisted $K$ source code files linked to those bug reports. Following this shortlisting, we employ the XGBoost model to predict the presence of a relevant file within the top $K$ files. If at least one relevant file exists, we proceed to the next step by passing the bug reports and filtered source code files.

\subsection{Formulation of RLocator}
\label{sec:formulation_of_RLocator}
In the previous step, we have pre-processed the dataset for training the reinforcement learning model. We shortlist $k$ most relevant files for each bug report. After that, we identify the bug reports for which there will be no relevant files in top $k$ files and filter out those bug reports. Finally, we pass the top $k$ relevant files to RLocator. In this section, we explain how RLocator employs Reinforcement Learning for its bug localization.
Our approach is grounded in the belief that each bug report contains specific indicators, such as terms and keywords, aiding developers in pinpointing problematic source code files. For example, in Java code, we can get a nested exception (the exception that leads to another exception, and another; an example is available in the online Appendix~\cite{replication}). A developer can identify the root exception and which method call (or any other code block) caused that. After that, they can go for the implementation of that method in the source code files. The process indicates that developers can identify and use the important information from a bug report. Following prior studies~\cite{Bagherzadeh2022, Wan2018}, we formulate RLocator into a Markov Decision Process (MDP) by dividing the ranking problem into a sequence of decision-making steps.
% For using the RL algorithms in RLocator, we have chosen to formulate the bug localization process into a Markov Decision Process (MDP). We have transformed the ranking problem into an MDP by dividing the ranking problem into a sequence of decision-making steps.
A general RL model can be represented by a tuple $\langle S, A, \tau, \Re, \pi\rangle$, which is composed of states, actions, transition, reward, and policy, respectively. 
Figure~\ref{fig:bl_as_mdp} shows an overview of our RLocator approach.
Next, we describe the formulation steps of each component of RLocator.\\
%  \begin{figure}[ht!]
%     \centering   
%     \includegraphics[scale=0.22]{Figures/code.png}
%     \includegraphics[scale=0.22]{Figures/stack.png}
%     \caption{Example code creating nested exception}
%     \label{fig:code}
% \end{figure}
\textbf{States:} $S$ is the set of states. The RL model moves from one state to another state until it reaches the end state. 
    To form the states of the MDP, we apply the following steps:
    \par \textbf{Input:} 
        % As shown in Figure~\ref{fig:bl_as_mdp}, 
        % the input of our MDP comprises two components, a bug report, and relevant source code files (documents) of a project repository.
        % In the next step, we identify $k$ relevant files (to the bug report) using ElasticSearch.
        % we show the bug report and the source codes associated with the bug report as input of our MDP.\par
        % Software projects maintain a repository for their bugs or issues (e.g., Jira, Github, Bugzilla). 
        % The first component, the bug report, can be retrieved from those issue repositories.
        % We use the bug report to obtain the second component (i.e., source code) by identifying the project version affected by the bug.         Typically, each bug report is associated with a version or commit SHA of the project repository. 
        % % To create the second input component (i.e., source code), we must identify the version affected by this particular bug from the bug report.
        % After identifying the buggy version, we collect all source code files from the specific version of the code repository. The source code files are indexed in the ELasticSearch, and then we query the index using a bug report. From ElasticSearch, we receive the top k relevant source code files. As shown in Figure~\ref{fig:bl_as_mdp}, 
        
        The input of our MDP comprises a bug report and the top $K$ relevant source code files from a project repository. We use CodeBERT~\cite{feng2020}, a transformer-based model, to convert text into embeddings, representing the text in a multi-dimensional space. CodeBERT is chosen for its ability to handle long contexts, making it suitable for long source code files where methods may be declared far from their usage. Unlike Word2Vec, which generates static embeddings for words, CodeBERT generates dynamic embeddings for sequences, capturing context during inference. This is crucial in source code files where variable use depends on scope.
        
        CodeBERT, trained on natural language and programming language pairs, handles both programming and natural languages. Its self-attention mechanism assesses the significance of individual terms, helping link bug reports to relevant source code files. For example, in a Java nested exception, developers can identify the main exception and pinpoint the responsible code block. RLocator relies on this self-attention mechanism to identify and leverage these informative cues effectively.
        
        In our approach, as shown in Figure~\ref{fig:bl_as_mdp}, the embedding model processes bug reports and source code files, generating embeddings for the source code files $F_1, F_2, ..., F_k$, and the bug report $R$.

        \par \textbf{Concatenation:} After we obtain the embedding for the source codes and the bug report, we concatenate them. As prior studies~\cite{He2017, Zhang2014} suggest combining distinct sets of features through concatenation and processing them with a linear layer enables effective interaction among the features. Furthermore, feature interaction is fundamental in determining similarity~\cite{Zhu2020, Khattab2020}. Thus, with the goal of calculating the similarity between a bug report and a source code file pair, we concatenate their embedding.
        Given our example in Figure~\ref{fig:bl_as_mdp}, we concatenate the embedding of $F_1, F_2, ...$, and $F_k$ with the embedding of bug report $R$ independently. 
       This step leads us to obtain the corresponding concatenated embedding $E_1, E_2, ...$, and $E_k$, as shown in Figure~\ref{fig:bl_as_mdp}.
 
    Note that each state of the MDP comprises two lists: a candidate list and a ranked list.
    The candidate list contains the concatenated list of embedding. As shown in our example in Figure~\ref{fig:bl_as_mdp}, the candidate list contains $E_1, E_2, ...$, and $E_k$.
    In the candidate list, source code embeddings (code files and bug reports embedding concatenated together) are ranked randomly. 
    The other list is the ranked list of source code files based on their relevance to the bug report $R$. 
    Initially (at $State_1$), the candidate list is full, and the ranked list is empty.
    In each state transition, the model moves one embedding from the candidate list to the ranked list based on their probability of being responsible for a bug. In the final state, the ranked list will be full, and the candidate list will be empty. 
    We describe the process of selecting and ranking a file in detail in the next step.
    
\noindent \textbf{Actions:} We define  \emph{Actions} in our MDP as selecting a file from the candidate list and moving it to the ranked list. Suppose at the timestep $t$; the RL model picks the embedding $E_1$, then the rank of that particular file will be $t$. In Figure~\ref{fig:bl_as_mdp} at the timestamp $1$, the model picks concatenated embedding of file $F_2$. Thus, the rank of $F_2$ will be $1$. As in each timestamp, we are moving one file from the candidate list to the ranked list; the total number of files will be equal to the number of states and the number of actions.
    For identifying the potentially best action at any timestamp $t$, we use a deep learning (DL) model (indicated as \emph{Ranking Model} in Figure~\ref{fig:bl_as_mdp}), which is composed of a Convolutional Neural Network (CNN) followed by a Long Short-Term Memory (LSTM)~\cite{Hochreiter1997}. Following~\cite{Pang2017, Huo2021, Huo2016}, we use CNN to establish the connection between source code files and bug reports and extract relevant features. As mentioned earlier, developers acquire the ability to recognize cues and subsequently employ them to establish the association between source code files and bug reports. The CNN facilitates the second stage of bug localization, which involves extracting important features. The input of the CNN is the concatenated embedding of both bug reports and each source code file, and the output of CNN is extracted features from the combined embedding of bug reports and source code files. The features are later used to calculate relevance.

    On the other hand, LSTM~\cite{Hausknecht2015}  intends to make the model aware of a restriction, which we call \emph{state awareness}. That is, in each timestamp, the model is allowed to pick the potentially best embedding that has not been picked yet, i.e., if a file is selected at $State_i$, it cannot be selected again in a later state (i.e., $State_{i+j};j\ge1$). The LSTM retains the state and aids the RL agent in choosing a subsequent action that does not conflict with prior actions.
    Thus, following previous studies~\cite{Matthew2015, Bello2016}, we use an LSTM to make the model aware of previous actions. The LSTM takes a set of feature vectors as input and outputs the id of the source code file most suitable for the current state.
    
    % However, we have found that our models' performance will remain almost the same even if we include that 15\% of data in the test dataset. We will discuss more about the performance lower limit in Section~\ref{sec:results}.
    
    % To do so, we have used \emph{action masking} approach to meet the requirement that the agent is not allowed to take a particular action more than once. The masking has been achieved by by multiplying the softmax probability of already used actions with zero.
\noindent \textbf{Transition:} $\tau$(S, A) is a function $\tau$ : S × A → S which maps a state $s_t$ into a new state $s_{t+1}$ in response to the selected action $a_t$. Choosing an action $a_t$ means removing a file from the candidate list and placing it in the ranked list.
    
\noindent \textbf{Reward:} A reward is a value provided to the RL agent as feedback on their action. 
    We refer to a reward received from one action as \emph{return}.
    The RL technique signals the agent about the appropriate action in each step through the \textit{Reward
Function}, which can be modeled using the retrieval metrics.
Thus, the RL agent can learn to
optimize the retrieval metrics through the reward function.
% , which we use to evaluate the agent.
We consider two important factors in the ranking evaluation: the position of the relevant files and the distance between relevant files in the ranked list of embedding. 
We incorporated both factors in designing the reward function shown below.
% We choose the following function as a reward:
\begin{multline}
            \Re(S,A) = \frac{ M * file\; relevance}{\log_2 (t+1) * distance(s)};if \;A \;is \\an \;action \;that \;has \;not \; been \; selected\; before \\
            \label{eq:reward_1}
        \end{multline}
        \begin{equation}
            \Re(S,A) = -\log_2 (t + 1) ; otherwise
            \label{eq:reward_2}
        \end{equation}
    
\begin{multline}
        distance (S) = Avg. (Distance\; between\;\; currently\\ picked\; subsequent\; related\; files)
\end{multline}

    \begin{figure}[tb!]
        \centering
        \includegraphics[scale=0.16]{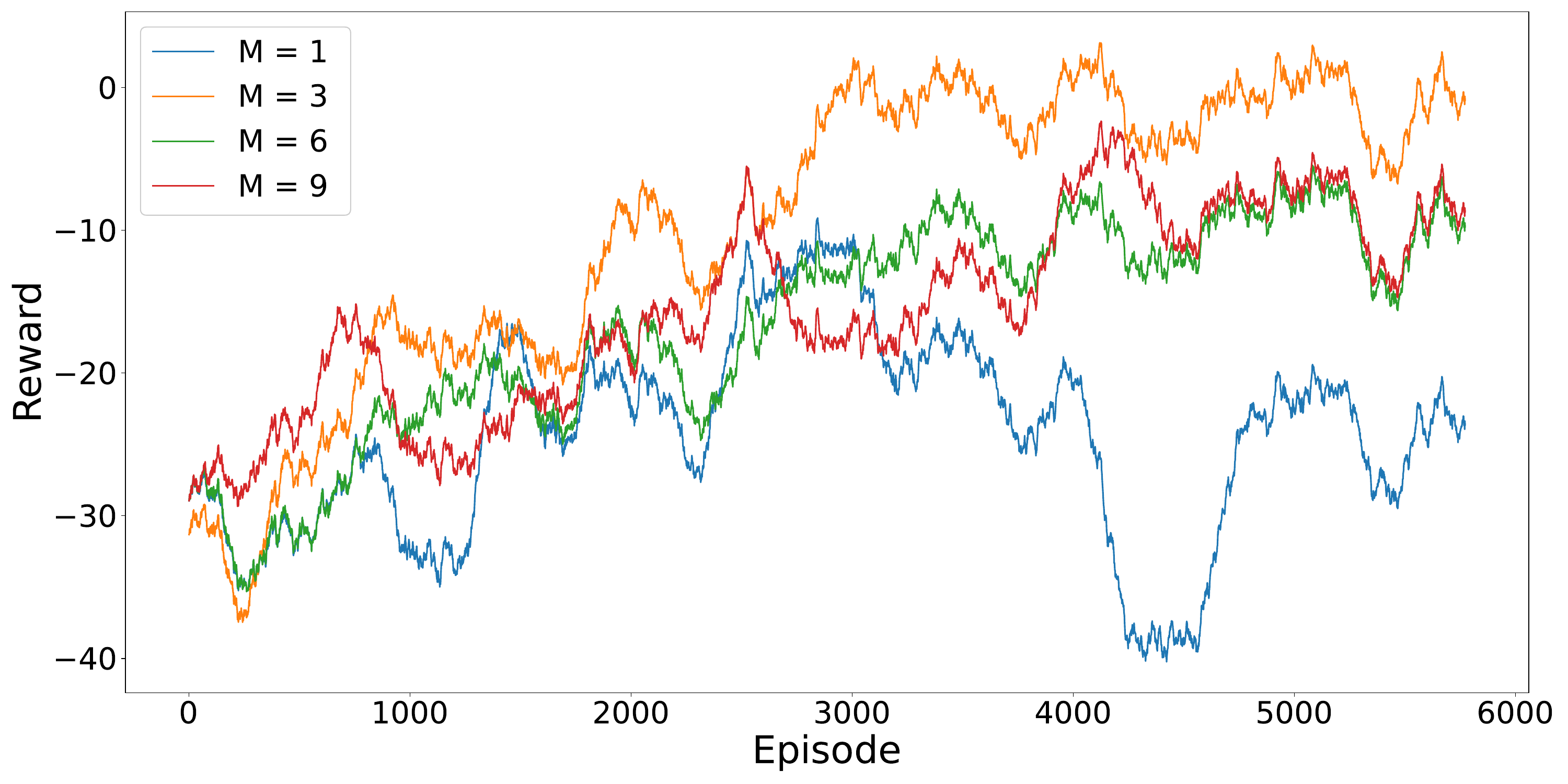}
        \caption{Effect of M in the reward-episode graph.}
        \label{fig:effect_m}
    \end{figure}
    
    % \begin{figure}[tb!]
    %     \centering
    %     \includegraphics[scale=0.16]{Figures/AspectJMultiplicator.eps}
    %     \caption{Effect of M in the reward-episode graph.}
    %     \label{fig:effect_m}
    % \end{figure}
    
    \noindent In Equations~\ref{eq:reward_1} and ~\ref{eq:reward_2}, $t$ is the timestamp, $S$ is State and $A$ is Action. Mean reciprocal rank (MRR) measures the average reciprocal rank of all the relevant files. In Equation~\ref{eq:reward_1}, $\frac{file\; relevance}{\log_2 (t+1)}$ represents the MRR. The use of a logarithmic function in the equation is motivated by previous studies~\cite{Haarnoja2018, Wang2019}, which  found that it leads to a stable loss. When the relevant files are ranked higher, the average precision tends to be higher. To encourage the reinforcement learning system to rank relevant files higher, we introduce a punishment mechanism if there is a greater distance between two relevant files. By imposing this punishment on the agent, we incentivize it to prioritize relevant files in higher ranks, which in turn contributes to the Mean Average Precision (MAP).
    
    We illustrate the reward functions with an example below. Presuming that the process reaches State $S_6$ and the currently picked concatenated embeddings are $E_1, E_2, E_3, E_4, E_5, E_6$ and their relevancy to the bug report is $\langle 0, 0, 1, 0, 1, 1\rangle$. This means that these embeddings (or files) ranked in the $3^{rd}$, $5^{th}$, and $6^{th}$ positions are relevant to the bug report. 
    The position of the relevant files are $\langle 3, 5, 6 \rangle$, and the distance between them is  $\langle 1, 0 \rangle$. Hence,  $distance(S_6) = Avg. \langle 1, 0 \rangle = 0.5$.
    If the agent picks a new relevant file, we reward the agent $M$ times the reciprocal rank of the file divided by the distance between the already picked related files. In our example, the last picked file, $E_6$'s relevancy is $1$. Thus, we have the following values for Equation~\ref{eq:reward_1}: $distance(S_6) = 0.5$; $\log_2 (6 + 1) = 2.8074$; $file\; relevance = 1$. 
    Note that $M$ is a hyper-parameter. We find that three as the value of M results in the highest reward for our RL model.
    We identify the best value for $M$ by experimenting with different values (1, 3, 6, and 9). Figure~\ref{fig:effect_m} shows the resulting reward-episode graph using different values of $M$.
    Hence, given $M=3$, the value of  the reward function will be $\Re(S,A) = \frac{3 * 1}{2.8074 * 0.5} = 2.14$. The reward can vary between $M$ to $\sim$ 0. A higher value of the reward function indicates a better action of the model. Finally, in the case of optimal ranking, the $distance(S)$ will be zero. We handle this case by assigning a value of $1$ for $distance(S)$.
    Even though we are using MRR and MAP as optimization goals we do not require labeled data. Instead, it learns from developers' feedback. It presents a limited set of files to the developer, seeking their feedback. If a developer deems a specific file as relevant, they can click on it. This click feedback signifies the file's relevance within the set. RLocator leverages this input at Equation~\ref{eq:reward_1} and learns the process of bug localization. Incorporating developers' feedback may cause some inconvenience. However, all machine learning models are prone to data drift~\cite{Rabinovich_2023, Islam_2021,Aljedaani_2018}, where initial training data no longer matches current data, leading to declining performance. RLocator addresses this by continuously updating its learning based on developers' feedback.\\
We limit the number of actions to 31 in RLocator. Since the number of states equals the number of actions, we also limit the number of states to 31. The prediction space of a reinforcement learning agent cannot be variable, and the number of source code files is variable. Thus, we must fix the number of actions, $k$, to a manageable number that fits in memory. We use an Nvidia V100 16 GB GPU and found that with more than 31 actions, training scripts fail due to out-of-memory errors. Therefore, we set $K=31$ to keep the state size under control. As mentioned in Section~\ref{sec: identification_of_unrankable_data}, we select the top 31 relevant source code files from ES and pass them to RLocator, ensuring the number of states and files remains the same.
\subsection{Developers' Workflow}
 \begin{figure}[ht!]
    \centering   
    \fbox{
        \includegraphics[scale=0.045]{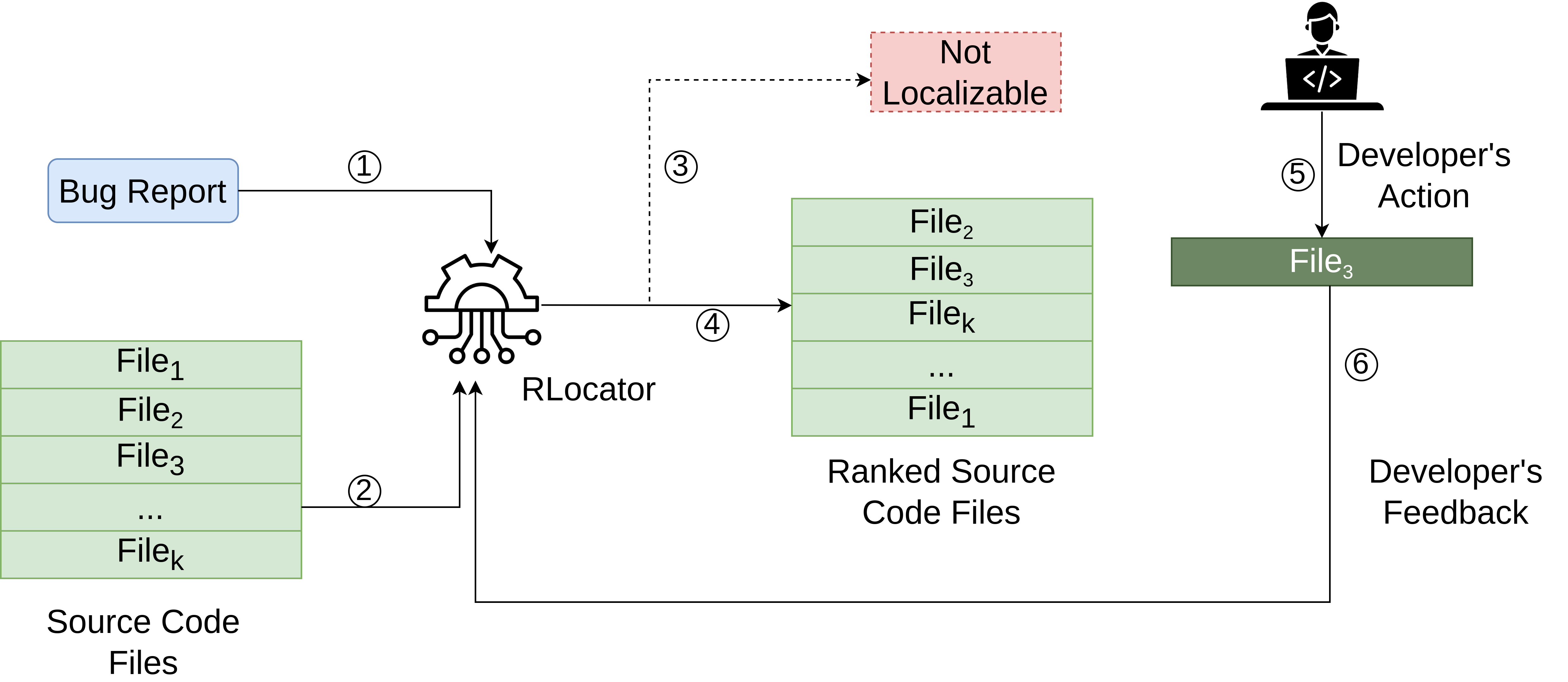}
    }
    \caption{Developer interaction flow.}
    \label{fig:interaction_flow}
\end{figure}

Figure~\ref{fig:interaction_flow} illustrates the interaction flow of developers using RLocator, which is represented as a central black box in the diagram. Details of RLocator are presented in Figure~\ref{fig:bl_as_mdp}. The process begins with RLocator receiving two primary inputs: a bug report and all the source code files, labeled as 1 and 2, respectively, in the figure. After processing these inputs, RLocator outputs a ranked list of 31 source code files, indicated by step \#4. Developers then review this list to identify and select files that may contain bugs; an example is shown when a developer selects $File_3$, noted as step \#5. This selection serves as feedback to RLocator, marked by step \#6, aiding in refining its bug localization strategy. The feedback from developers is expressed as a binary value: files that developers open are marked with a 1, and all other files are marked with a 0. Additionally, the system can also indicate its inability to localize a bug for a given report, as shown by step \#3. This ongoing loop allows RLocator to stay updated with changes in techniques and patterns, enhancing its bug localization performance.

% Developers will follow a similar training path outlined in Figure~\ref{fig:bl_as_mdp}. The distinction lies in the training dataset: multiple bug reports and source code files were used, while developers will use a single bug report and its associated source code files. Following this, the source code file will undergo ES-based filtration and use the XGBoost model. RLocator generates embeddings for the bug report and source code file using the CodeBERT model based on the classifier's prediction. Subsequently, RLocator will present a file set to the developer based on ranking models' predictions, allowing for developer feedback. Please note that, during training, developers' feedback was simulated, but in a real-world scenario, RLocator will directly accept feedback from developers. Based on the developers' feedback RLocator will correct itself and will show a new set of files to the developer if it had a wrong prediction.

\section{Dataset and Evaluation Measures}
\label{sec:dataset}
In this section, we discuss the dataset used to train and evaluate our model (section ~\ref{dataset}).
Then, we present the evaluation metrics we use for evaluation (section~\ref{sec:evaluation}).

\subsection{Dataset}
\label{dataset}
In our experiment, we evaluate our approach on six real-world open-source projects~\cite{Ye2014}, which are commonly used benchmark datasets in bug localization studies~\cite{Liang2022,Zhang2020}. Prior work has shown that this dataset has the lowest number of false positives and negatives compared to other datasets~\cite{Lee2018, Dit2011, Zhou2012, Moreno2015, Sisman2013}. Following previous studies, we train our RLocator model separately for each of the six Apache projects (AspectJ, Birt, Eclipse Platform UI, JDT, SWT, Tomcat). Table~\ref{table:data_stat} shows descriptive statistics on the datasets.

The dataset contains metadata such as bug ID, description, report timestamp, commit SHA of the fixing commit and buggy source code file paths. Each bug report is associated with a commit SHA/version, and we use a multiple version set matching approach to exclusively utilize the source code files linked to each specific report~\cite{Lee2018}. This approach closely resembles the bug localization process done by developers and reduces noise in the dataset, improving tool performance.

We identify the version containing the bug from the commit SHA and collect all relevant source code files from that version, excluding the bug-fixing code. This ensures our bug localization system closely mimics real-world scenarios.

For training and testing, we use 91\% of the data, sorting the dataset by the date of bug reports and splitting it 60:40 for training and testing, respectively. Unlike previous studies~\cite{Xiao2019} that used a 60:20:20 split, we repurpose validation data for testing to shorten the training duration.

\begin{table}[tb!]
\centering
\caption{Dataset statistics.}
% \resizebox{0.6\columnwidth}{!}{%
\begin{tabular}{@{}crr@{}}
\toprule
\textbf{Project} & \textbf{\begin{tabular}[c]{@{}r@{}}\# of Bug\\ Reports\end{tabular}} & \textbf{\begin{tabular}[c]{@{}r@{}}Avg. \# of Buggy\\ Files per Bug\end{tabular}} \\ \midrule
AspectJ          & 593                                                                 & 4.0                                                                              \\ \midrule
Birt             & 6,182                                                               & 3.8                                                                              \\ \midrule
Eclipse UI       & 6,495                                                               & 2.7                                                                              \\ \midrule
JDT              & 6,274                                                               & 2.6                                                                              \\ \midrule
SWT              & 4,151                                                               & 2.1                                                                              \\ \midrule
Tomcat           & 1,056                                                               & 2.4                                                                              \\ \bottomrule
\end{tabular}
\label{table:data_stat}

\end{table}

\subsection{ Evaluation Measures}
\label{sec:evaluation}
The dataset proposed by Ye et al.~\cite{Ye2014} provides ground truth associated with each bug report. 
The ground truth contains the path of the file in the project repository that has been modified to fix a particular bug.
To evaluate RLocator performance, we use the ground truth and analyze
the experimental results based on three criteria, which are widely adopted in bug localization studies~\cite{Zhou2012,Liang2019, Zhang2020,Zhu2022,Ciborowska2022}.
% Using the ground truth we can evaluate the performance of our proposed model. 
% Following previous studies, we have used Mean Reciprocal Rank (MRR) and Mean Average Precision (MAP) as evaluation metrics
\begin{itemize}
    
    \item \textbf{Mean Reciprocal Rank (MRR):} To identify the average rank of the relevant file in the retrieved files set, we adopted the Mean Reciprocal Rank. MRR is the average reciprocal rank of the source code files for all the bug reports. We present the equation for calculating MRR below, where $A$ is the set of bug reports.
    \begin{equation}
        MRR= \frac{1}{|A|}\sum_{A}\frac{1}{Least\; rank\; of\; the\; relevant\; files}
    \end{equation}
    Suppose we have two bug reports, $report_1$ and $report_2$. For each bug report, the bug localization model will rank six files. For $report_1$ the ground truth of the retrieved files are $[0, 0, 1, 0, 1, 0]$ and for $report_2$ the ground truth of the retrieved files are $[1, 0, 0, 0, 0, 1]$. In this case, the least rank of relevant files is 3 and 1, respectively, for $report_1$ and $report_2$. Now, the $MRR = \frac{1}{2}(\frac{1}{3} + \frac{1}{1}) = 0.67$\\
    
    \item \textbf{Mean Average Precision (MAP):} To consider the case where a bug is associated with multiple source code files, we adopted Mean Average Precision. It provides a measure of the quality of the retrieval~\cite{Zhou2012, Schwarz1981}. MRR considers only the best rank of relevant files; on the contrary, MAP considers the rank of all the relevant files in the retrieved files list. Thus, MAP is more descriptive and unbiased than MRR. Precision means how noisy the retrieval is. If we calculate the precision on the first two retrieved files, we will get precision@2. For calculating average precision, we have to figure precision@1, precision@2,... precision@k, and then we have to average the precision at different points. After calculating the average precision for each bug report, we have to find the mean of the average precision to calculate the MAP.
    \begin{equation}
        MAP = \frac{1}{|A|}\sum_{A}{AvgPrecision(Report_i)}
    \end{equation}
    
    We show the MAP calculation for the previous example of two bug reports. The Average precision for $report_1$ and $report_2$ will be 0.37 and 0.67. So, the $MAP = \frac{1}{2}(0.36 + 0.67) = 0.52$

    \item \textbf{Top K:}
    For fare comparison with prior studies~\cite{Huo2016, Huo2021} and to present a straightforward understanding of performance we calculate Top K. Top K measures the overall ranking performance of the bug localization model. It indicates the percentage of bug reports for which at least one buggy source  appears among the top K positions in the ranked list generated by the bug localization tool. Following previous studies (e.g., ~\cite{Huo2016, Huo2021}), we consider three values of K: 1, 5, and 10.
    
\end{itemize}

\section{RLocator Performance}
\label{sec:results}

%  \begin{figure*}[ht!]
%     \centering
%     \includegraphics[scale=0.45]{Figures/AspectJ_R_E.eps}
%     \caption{Reward-Episode graph for training on AspectJ bug reports}
%     \label{fig:re_graph}
% \end{figure*}
We evaluate RLocator on the hold-out dataset using the metrics described in Section~\ref{sec:evaluation}. 
As there has been no RL-based bug localization tool, we compare RLocator with three state-of-the-art bug localization tools: BugLocator, FLIM, and BL-GAN.

% including DeepLoc, CAST, KGBugLocator, BL-GAN, FBL-BERT, FLIM, and baseline tool BugLocator. Among these methods, BugLocator is an information-retrieval-based method. DeepLoc, CAST, KGBugLocator, BL-GAN, FBL-BERT, and 

A short description of the approaches is presented below.
% \usepackage{multirow}% Please add the following required packages to your document preamble:
% \usepackage{multirow}

\begin{table*}[]
\centering
\caption{RLocator performance.}
\begin{tabular}{@{}llcccccccccc@{}}
\toprule
\multicolumn{1}{c}{\multirow{2}{*}{\textbf{Project}}} & \multicolumn{1}{c}{\multirow{2}{*}{\textbf{Model}}} & \multicolumn{2}{c}{\textbf{Top 1}}                  & \multicolumn{2}{c}{\textbf{Top 5}}                  & \multicolumn{2}{c}{\textbf{Top 10}}                 & \multicolumn{2}{c}{\textbf{MAP}}                    & \multicolumn{2}{c}{\textbf{MRR}}                    \\ \cmidrule(l){3-12} 
\multicolumn{1}{c}{}                                  & \multicolumn{1}{c}{}                                & \textbf{91\%}            & \textbf{100\%}           & \textbf{91\%}            & \textbf{100\%}           & \textbf{91\%}            & \textbf{100\%}           & \textbf{91\%}            & \textbf{100\%}           & \textbf{91\%}            & \textbf{100\%}           \\ \midrule
\multirow{5}{*}{AspectJ}                              & RLocator                                            & 0.46                     & 0.40                     & 0.69                     & 0.63                     & 0.75                     & 0.70                     & 0.56                     & 0.46                     & 0.59                     & 0.50                     \\ \cmidrule(l){2-12} 
                                                      & BugLocator                                          & 0.36                     & 0.28                     & 0.50                     & 0.45                     & 0.56                     & 0.51                     & 0.33                     & 0.31                     & 0.49                     & 0.48                     \\ \cmidrule(l){2-12} 
                                                      & FLIM                                                & 0.51                     & 0.36                     & 0.65                     & 0.60                     & 0.72                     & 0.67                     & 0.41                     & 0.35                     & 0.47                     & 0.45                     \\ \cmidrule(l){2-12} 
                                                      & CodeBERT                                            & 0.4                      & 0.35                     & 0.59                     & 0.55                     & 0.65                     & 0.61                     & 0.49                     & 0.39                     & 0.51                     & 0.44                     \\ \cmidrule(l){2-12} 
                                                      & BL-GAN                                              & 0.41                     & 0.38                     & 0.6                      & 0.55                     & 0.71                     & 0.65                     & 0.33                     & 0.31                     & 0.42                     & 0.39                     \\ \midrule
\multirow{5}{*}{Birt}                                 & RLocator                                            & 0.65                     & 0.25                     & 0.46                     & 0.41                     & 0.53                     & 0.48                     & 0.47                     & 0.38                     & 0.49                     & 0.41                     \\ \cmidrule(l){2-12} 
                                                      & BugLocator                                          & 0.61                     & 0.15                     & 0.27                     & 0.21                     & 0.34                     & 0.29                     & 0.30                     & 0.30                     & 0.39                     & 0.38                     \\ \cmidrule(l){2-12} 
                                                      & FLIM                                                & 0.49                     & 0.18                     & 0.39                     & 0.34                     & 0.47                     & 0.42                     & 0.29                     & 0.25                     & 0.31                     & 0.28                     \\ \cmidrule(l){2-12} 
                                                      & CodeBERT                                            & 0.33                     & 0.22                     & 0.39                     & 0.35                     & 0.46                     & 0.43                     & 0.41                     & 0.33                     & 0.42                     & 0.35                     \\ \cmidrule(l){2-12} 
                                                      & BL-GAN                                              & 0.17                     & 0.16                     & 0.33                     & 0.3                      & 0.46                     & 0.42                     & 0.32                     & 0.29                     & 0.4                      & 0.37                     \\ \midrule
\multirow{5}{*}{Eclipse Platform UI}                  & RLocator                                            & 0.45                     & 0.37                     & 0.69                     & 0.63                     & 0.78                     & 0.73                     & 0.54                     & 0.42                     & 0.59                     & 0.50                     \\ \cmidrule(l){2-12} 
                                                      & BugLocator                                          & 0.45                     & 0.33                     & 0.54                     & 0.49                     & 0.63                     & 0.58                     & 0.29                     & 0.30                     & 0.38                     & 0.35                     \\ \cmidrule(l){2-12} 
                                                      & FLIM                                                & 0.48                     & 0.41                     & 0.72                     & 0.67                     & 0.80                     & 0.75                     & 0.51                     & 0.48                     & 0.52                     & 0.53                     \\ \cmidrule(l){2-12} 
                                                      & CodeBERT                                            & 0.39                     & 0.32                     & 0.6                      & 0.55                     & 0.68                     & 0.62                     & 0.47                     & 0.36                     & 0.52                     & 0.44                     \\ \cmidrule(l){2-12} 
                                                      & BL-GAN                                              & \multicolumn{1}{l}{0.34} & \multicolumn{1}{l}{0.31} & \multicolumn{1}{l}{0.53} & \multicolumn{1}{l}{0.49} & \multicolumn{1}{l}{0.66} & \multicolumn{1}{l}{0.61} & \multicolumn{1}{l}{0.32} & \multicolumn{1}{l}{0.3}  & \multicolumn{1}{l}{0.4}  & \multicolumn{1}{l}{0.36} \\ \midrule
\multirow{5}{*}{JDT}                                  & RLocator                                            & 0.44                     & 0.33                     & 0.67                     & 0.61                     & 0.78                     & 0.75                     & 0.51                     & 0.44                     & 0.53                     & 0.45                     \\ \cmidrule(l){2-12} 
                                                      & BugLocator                                          & 0.34                     & 0.21                     & 0.51                     & 0.45                     & 0.60                     & 0.55                     & 0.22                     & 0.20                     & 0.31                     & 0.28                     \\ \cmidrule(l){2-12} 
                                                      & FLIM                                                & 0.40                     & 0.35                     & 0.65                     & 0.60                     & 0.82                     & 0.77                     & 0.42                     & 0.41                     & 0.51                     & 0.49                     \\ \cmidrule(l){2-12} 
                                                      & CodeBERT                                            & 0.38                     & 0.29                     & 0.59                     & 0.54                     & 0.68                     & 0.66                     & 0.44                     & 0.38                     & 0.46                     & 0.39                     \\ \cmidrule(l){2-12} 
                                                      & BL-GAN                                              & 0.3                      & 0.27                     & 0.53                     & 0.48                     & 0.64                     & 0.59                     & 0.35                     & 0.32                     & 0.44                     & 0.41                     \\ \midrule
\multirow{5}{*}{SWT}                                  & RLocator                                            & 0.40                     & 0.30                     & 0.57                     & 0.51                     & 0.63                     & 0.58                     & 0.48                     & 0.42                     & 0.51                     & 0.44                     \\ \cmidrule(l){2-12} 
                                                      & BugLocator                                          & 0.37                     & 0.25                     & 0.50                     & 0.45                     & 0.56                     & 0.51                     & 0.42                     & 0.40                     & 0.46                     & 0.43                     \\ \cmidrule(l){2-12} 
                                                      & FLIM                                                & 0.51                     & 0.37                     & 0.70                     & 0.65                     & 0.83                     & 0.78                     & 0.43                     & 0.43                     & 0.48                     & 0.50                     \\ \cmidrule(l){2-12} 
                                                      & CodeBERT                                            & 0.34                     & 0.27                     & 0.5                      & 0.45                     & 0.54                     & 0.51                     & 0.42                     & 0.37                     & 0.45                     & 0.39                     \\ \cmidrule(l){2-12} 
                                                      & BL-GAN                                              & \multicolumn{1}{l}{0.31} & \multicolumn{1}{l}{0.29} & \multicolumn{1}{l}{0.53} & \multicolumn{1}{l}{0.48} & \multicolumn{1}{l}{0.6}  & \multicolumn{1}{l}{0.55} & \multicolumn{1}{l}{0.37} & \multicolumn{1}{l}{0.34} & \multicolumn{1}{l}{0.44} & \multicolumn{1}{l}{0.4}  \\ \midrule
\multirow{5}{*}{Tomcat}                               & RLocator                                            & 0.46                     & 0.39                     & 0.61                     & 0.55                     & 0.73                     & 0.68                     & 0.59                     & 0.47                     & 0.62                     & 0.51                     \\ \cmidrule(l){2-12} 
                                                      & BugLocator                                          & 0.40                     & 0.29                     & 0.43                     & 0.38                     & 0.55                     & 0.50                     & 0.31                     & 0.27                     & 0.37                     & 0.35                     \\ \cmidrule(l){2-12} 
                                                      & FLIM                                                & 0.51                     & 0.42                     & 0.70                     & 0.65                     & 0.76                     & 0.71                     & 0.52                     & 0.47                     & 0.59                     & 0.60                     \\ \cmidrule(l){2-12} 
                                                      & CodeBERT                                            & 0.39                     & 0.34                     & 0.53                     & 0.49                     & 0.62                     & 0.6                      & 0.51                     & 0.41                     & 0.53                     & 0.44                     \\ \cmidrule(l){2-12} 
                                                      & BL-GAN                                              & \multicolumn{1}{l}{0.38} & \multicolumn{1}{l}{0.35} & \multicolumn{1}{l}{0.61} & \multicolumn{1}{l}{0.55} & \multicolumn{1}{l}{0.65} & \multicolumn{1}{l}{0.61} & \multicolumn{1}{l}{0.43} & \multicolumn{1}{l}{0.4}  & \multicolumn{1}{l}{0.55} & \multicolumn{1}{l}{0.5}  \\ \bottomrule
\end{tabular}
\label{table:result}
\end{table*}

\begin{itemize}
    \item BugLocator~\cite{Zhou2012}: an IR-based tool that utilizes a vector space model to identify the potentially responsible source code files by estimating the similarity between source code file and bug report.
    \item FLIM~\cite{Liang2022}: a deep-learning-based model that utilizes a large language model like CodeBERT.
    % \item CAST~\cite{Liang2019}:  a deep learning model that utilizes both the lexical and program semantics to identify the buggy files.
    % \item KGBugLocator~\cite{Zhang2020}: a graph embedding-based tool that utilizes an attention mechanism to localize bugs. 
    \item BL-GAN~\cite{Zhu2022}: uses generative adversarial strategy to train an attention-based transformer model. 
    
\end{itemize}

We use the original implementations to assess the performance of BugLocator~\cite{Zhou2012} and FLIM~\cite{Liang2022}. Additionally, we fine-tune a CodeBERT~\cite{feng2020} model as a baseline to demonstrate the benefits of using reinforcement learning. For tools like CAST~\cite{Liang2019}, KGBugLocator~\cite{Zhang2020}, and BL-GAN~\cite{Zhu2022}, which lack replication packages, we refer to their respective studies. These studies show that KGBugLocator outperforms CAST, and BL-GAN outperforms KGBugLocator. Consequently, we replicate BL-GAN based on its study descriptions.

Regarding FBL-BERT~\cite{Ciborowska2022}, a recent technique, we do not compare it with RLocator. This is because FBL-BERT performs bug localization at the changeset level, and applying it to our file-level dataset would disadvantage FBL-BERT, as it is designed for shorter documents. Therefore, comparing it with RLocator would be unfair.

Furthermore, other studies, such as DeepLoc~\cite{Xiao2019}, bjXnet~\cite{Han2023}, CAST~\cite{Liang2019}, KGBugLocator~\cite{Zhang2020}, and Cheng et al.~\cite{Cheng2020}, also propose deep learning-based approaches but do not provide replication packages. Although these studies evaluate similar projects, the lack of available code or pre-trained models prevents further comparison. However, to ensure comprehensive information, we include a table in our online appendix~\cite{replication} displaying their performance alongside RLocator.

\subsection{Retrieval performance}
\label{sec:retrieval_results}
% In section~\ref{sec:formulation_of_RLocator},
% As previously mentioned, we utilize $k$ (=31) relevant files in RLocator. 
% Note that with $k$=31 and using ES as we describe in Section~\ref{sec: identification_of_unrankable_data}, we can rerank files using RLocator for 91\% of the bug reports in our dataset.
% Table~\ref{table:result} presents the performance of RLocator along with the performance of the examined studies considering 91\% of the data and 100\% of the data. 
% Note that RLocator is not designed to utilize 100\% data since it cannot rerank the files for any bug report for which there are no relevant files among the top $k$ files. 
% Thus, for the cases where RLocator is not applicable (no relevant files identified by the XGBoost model), we estimate the performance of RLocator. For estimation, we assume the contribution of the case is zero. For example, if we have three bug reports where for the $1^{st}$ bug report, the rank of the relevant files is three and five, $2^{nd}$ bug report, the rank of the files is five and seven and the $3^{rd}$ bug report the XGBoost model predicts that there will be no relevant file in top $K$ files. The MRR for this scenario will be $\frac{1}{3} (\frac{1}{2} + \frac{1}{5} + 0) = 0.23$. As we are using zero instead of reciprocal ranks when calculating Mean Reciprocal Rank (MRR)  (or MAP), we can say that estimated performance is the worst possible performance or performance lower bound of RLocator. We choose such an overly conservative approach to avoid overestimating the effectiveness of our technique. 
We use $k$=31 relevant files in RLocator, allowing us to rerank files for 91\% of the bug reports. Table~\ref{table:result} shows RLocator's performance on 91\% and 100\% of the data. RLocator is not designed for 100\% data as it cannot rerank files if no relevant files are in the top $k$ files. For such cases, we estimate performance assuming zero contribution, providing a lower bound for RLocator's effectiveness. This conservative approach ensures we do not overestimate the technique's effectiveness.
\textbf{Table~\ref{table:result} showcases that RLocator achieves better performance than BugLocator and FLIM in both MRR and MAP across all studied projects when using the 91\% data.}
On 91\% data, RLocator outperforms FLIM by 5.56-38.3\% in MAP and 3.77-36.73\% in MRR. 
Regarding Top K, the performance improvement is up to 23.68\%, 15.22\%, and 11.32\% in terms of Top 1, Top 5, and Top 10, respectively. 
Compared to BugLocator, RLocator achieves performance improvement of 12.5 - 56.86\% and 9.8\% - 41.51\%, in terms of MAP and MRR, respectively. Regarding Top K, the performance improvement is up to 26.32\%, 41.3\%, and 35.85\% in terms of Top 1, Top 5, and Top 10, respectively. The results indicate that RLocator consistently outperforms BL-GAN in 91\% settings across all metrics. Specifically, in the TopK measurements, RLocator's performance exceeded that of BL-GAN, with improvements ranging from 55.26\% to 3.33\%. Additionally, RLocator achieved performance gains of 40.74\% in MAP and 32.2\% in MRR, respectively.

The results point out that RLocator outperforms BL-GAN across all the metrics in 91\% settings. Specifically, in TopK, RLocator achieved better performance than BL-GAN, ranging from 3.33\% to 55.26\%. The performance gain is 40.74\% and 32.2\% for MAP and MRR, respectively.

Compared to the CodeBERT model trained as a classifier (CodeBERT), RLocator achieves better performance across all the metrics. CodeBERT model archives consistently lower performance across all the metrics. The performance drops up to 17.65\%, 15.63\%, 17.95\%, 17.14\%, and 16.67\% for Top1, Top5, Top 10, MAP, and MRR, respectively.

% \tbl{Note that we could not conduct the same analysis for CAST and KGBugLocator due to the lack of a replication package. Thus, we are not discussing the performance of CAST and KGBugLocator in this study. However, they are presented in Table~\ref{table:result} for a complete understanding.}
% \mei{Update the percentages below and check that my findings are accurate.}

When we consider 100\% of the data, RLocator has better MAP results than FLIM in three out of the six projects (AspectJ, Birt, and JDT) by 6.82-34.21\%, equal to FLIM in one project (Tomcat) and worse than FLIM in 2 projects (Eclipse Platform UI and SWT) by 2-14\%. \tbl{The underperformance of RLocator in the Eclipse Platform UI and SWT projects can be linked to the poor and inconsistent quality of bug reports, which creates a significant lexical gap between the reports and the source code. By applying the IMaChecker~\cite{Soltani2020} approach, we discovered that AspectJ reports are of the highest quality, whereas those for Eclipse Platform UI, SWT, and Tomcat rank among the lowest. For a detailed analysis of bug report quality, please refer to the online Appendix~\cite{replication}.}
In terms of MRR, RLocator is better than FLIM in 2 projects (AspectJ and Birt) by 10-31.71\% and worse than FLIM in the remaining four projects (Eclipse platform UI, JDT, SWT, and Tomcat) by 6-18\%. 
In terms of Top K, RLocator ranks 4.29-12.5\% more bugs in the top 10 positions than FLIM in two projects. On the other hand, in the rest of the four projects, FLIM ranks more bugs in the top 10 positions, ranging between 2.74 - 34.48\%. When comparing RLocator with BugLocator for the 100\% data along MAP, we find that the RLocator is better in five of the six projects and similar in just the Tomcat project. 
With respect to MRR, RLocator is better than BugLocator in all six projects. In terms of Top K, RLocator ranks more bugs than BugLocator in the top 10 position, where the improvement ranges between 12.07 -39.58\%.
The results demonstrate that RLocator consistently surpasses BL-GAN in all metrics in the 100\% setting. Specifically, in the TopK metric, RLocator's performance was better than BL-GAN, with improvements ranging from 3.33\% to 36\%. The performance enhancements for RLocator are 32\% in MAP and 1.96\% in MRR, respectively.

% \tbl{In comparison to CAST, RLocator has demonstrated varying levels of performance across different projects. Specifically, RLocator outperformed CAST in terms of MAP in one project, achieved similar results in another, and showed lower performance in a third project. It is worth noting that CAST did not analyze the Birt and Eclipse projects, so a direct performance comparison is not possible for those particular projects. When considering MRR, RLocator exhibited inferior performance in four projects, with performance degradation ranging from 2.1\% to 12\%. In the context of the top K rankings, RLocator ranked an equivalent number of bugs as CAST in one project (JDT) and fewer bugs, up to a maximum of 17\%, in the remaining three projects (AspectJ, SWT, Tomcat).}

% \tbl{Comparatively, when assessing RLocator against KGBuglocator, it achieved a superior MAP in two projects and a lower MAP in two projects. As for MRR, RLocator showed a drop in performance in four projects, with a decline ranging from 4.2\% to 17\%. Regarding top K rankings, RLocator ranked the same number of bugs as KGBuglocator in one project (JDT) and fewer bugs, up to a maximum of 30\%, in the other three projects (AspectJ, SWT, Tomcat).}

% \tbl{When compared to BL-GAN, RLocator demonstrated a better MAP in four projects. In terms of top K rankings, RLocator ranked more bugs within the top 10 for AspectJ, Eclipse, and JDT while ranking 8\% fewer bugs in the Tomcat project.}

It is important to note that 
% as explained in Section ~\ref{sec:evaluation}, 
MAP provides a more balanced view than MRR and top K since it accounts for all the files that are related to a bug report and not just one file. Additionally, in our technique, we optimize to give more accurate results for most of the bug reports than give less accurate results on average for all the bug reports. Thus, \textbf{by looking at the MAP data for the 91\%, we can see that RLocator performs better than the state-of-the-art techniques in all projects. Even if we consider 100\% of the data, RLocator is still better than other techniques in the majority of the projects.}
Only with 100\% of the data and when using MRR as the evaluation metric, RLocator does not perform better than the state-of-the-art in most projects.

RLocator performs the worst in the Birt project, with a performance drop of 10.47\% in MAP, 11.71\% in MRR, and 41.42\% in the top 10 compared to its average on 91\% of the data. Despite this, RLocator outperforms FLIM by 38.3\% in MAP, 36.73\% in MRR, and 11.32\% in the top 10. It also surpasses BugLocator by 36.17\% in MAP, 20.41\% in MRR, and 35.85\% in the top 10. Factors like bug report quality, amount of information in the bug report, and source code length may contribute to the performance drop. We measure helpful information in bug reports using a \emph{similarity} metric, which calculates the ratio of similar tokens between source code files and bug reports, indicating the potential usefulness of the report for bug localization. The metric is defined in  equation~\ref{eq:similarity}.
% Across all projects, we observe that RLocator performs the lowest in the Birt project. Interestingly, besides RLocator, 
% FLIM  also achieved the lowest performance (in both MRR and MAP). Compared to the average performance on 91\% data, RLocator's performance has dropped by nearly 10.47\%, 11.71\%, and 41.42\% in terms of MAP, MRR, and top 10, respectively. Nevertheless, compared with FLIM, the performance is 38.3\%, 36.73\%, and 11.32\% better for MAP, MRR, and the top 10,  respectively. 
% With respect to the BugLocator technique, the performance is 36.17\%, 20.41\%, and 35.85\% better in terms of MAP, MRR, and top 10, respectively. 
% % The performance drop can be one of the reasons behind the selective inclusion of projects in previous studies. 
% In fact, several factors can contribute to such performance drop~\cite{Mills2017} in the Birt project, e.g., the quality of bug reports, source code length, amount of information in the bug report,  etc. 
% As \emph{similarity} is one of the  important criteria for text retrieval-based bug localization systems~\cite{Moreno2013}, we estimate the amount of helpful information by calculating \emph{similarity} according to equation~\ref{eq:similarity}.
% The similarity metric measures the ratio of similar tokens between source code files and bug reports. High similarity indicates that the bug report contains much potentially helpful information to localize the bug.

\begin{equation}
    \label{eq:similarity}
    \hspace*{-1em}
        Similarity = \frac{Bug\ Report\ Tokens \cap File\ Tokens}{\#\ of\ Unique\ Tokens\ in\ Bug\ Report}
    \end{equation}
\tbl{The median similarity scores for the Birt, Eclipse Platform UI, and SWT projects are 0.29, 0.30, and 0.33, respectively, making them the lowest among the six projects. This observation suggests that the lower quality of bug reports (reflected in their similarity to source files) may contribute to the decreased performance of RLocator in these projects.}
% Thus, we believe that because of the bug reports' quality, the performance of RLocator dropped in the Birt project.\par

\begin{figure}[tb!]
    \centering
    \includegraphics[scale=0.16]{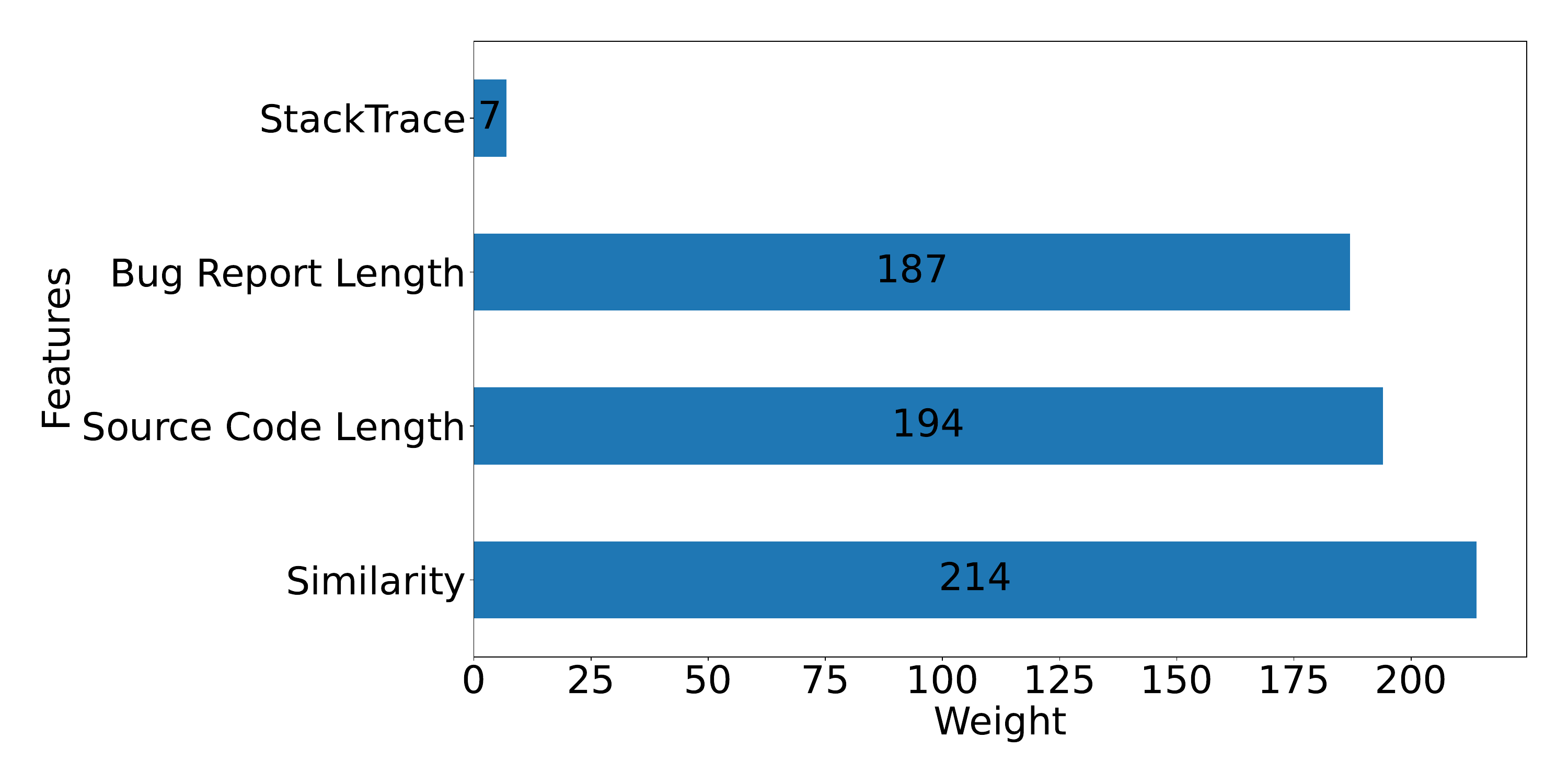}
    \caption{Feature importance of classifier model.}
    \label{fig:feature_importance}
\end{figure}
To effectively use RLocator in real-world scenarios, we employ an XGBoost model (Section~\ref{sec: identification_of_unrankable_data}) to filter out bug reports where relevant files do not appear in the top $K$ (=31) files. We then compute the importance of features listed in Table~\ref{table:feature_description} using XGBoost's built-in module. The importance score indicates each feature's contribution to the model, with higher values signifying greater importance. Figure~\ref{fig:feature_importance} shows that similarity is the most crucial feature, followed by source code length and Bbg report length. These findings highlight the importance of similarity in text-based search systems and suggest that high-quality bug reports can influence localization performance.

\subsection{Entropy Ablation Analysis: Impact of Entropy on RLocator performance}
\label{subsec: entropy}
We conduct an ablation study to gain insights into the significance of each component of RLocator. The main two components of RLocator are the ES-based shortlisting step and the reinforcement learning step.\\
In RL, we have used the A2C with entropy algorithm. Entropy refers to the unpredictability of an agent's actions. A low entropy indicates a predictable policy, while high entropy represents a more random and robust policy. An agent in RL will tend to repeat actions that previously resulted in positive rewards while learning the policy. The agent may become stuck in a local optimum due to exploiting learned actions instead of exploring new ones and finding a higher global optimum. This is where entropy comes useful: we can use entropy to encourage exploration and avoid getting stuck in local optima~\cite{Ahmed2019}. Because of this,  entropy in RL has become very popular in the design of RL approaches such as A2C~\cite{Jang2022}. In our proposed model (Section~\ref{sec:retrieval_results}), we use A2C with entropy to train the RLocator aiming to rank relevant files closer to each other.
As entropy is part of the reward, the gradient descent process will try to maximize the entropy. Entropy will increase if the model identifies different actions as the best in the same state.
However, those actions must select a relevant file; otherwise, the reward will be decreased. Thus, if there are multiple relevant files in a state, the A2C with entropy regularized model will assign almost the same probability in those actions (actions related to selecting those relevant files). This means that when the states are repeated, a different action will likely be selected each time. This probability assignment will lead to a higher MAP.

% \usepackage{multirow}
\setlength{\tabcolsep}{3pt}
\begin{table}[tb!]
\centering

\caption{RLocator performance with and without Entropy for A2C.}
\begin{tabular}{@{}ccrrrrr@{}}
\toprule
\textbf{Project}                                                               & \textbf{Model}   & \multicolumn{1}{c}{\textbf{Top 1}} & \multicolumn{1}{c}{\textbf{Top 5}} & \multicolumn{1}{c}{\textbf{Top 10}} & \multicolumn{1}{c}{\textbf{MAP}} & \multicolumn{1}{c}{\textbf{MRR}} \\ \midrule
\multirow{3}{*}{AspectJ}                                                       & ES               & 0.15                               & 0.20                               & 0.28                                & 0.23                             & 0.27                             \\ \cmidrule(l){2-7} 
                                                                               & A2C              & 0.27                               & 0.39                               & 0.48                                & 0.40                             & 0.52                             \\ \cmidrule(l){2-7} 
                                                                               & A2C with Entropy & 0.46                               & 0.69                               & 0.75                                & 0.56                             & 0.59                             \\ \midrule
\multirow{3}{*}{Birt}                                                          & ES               & 0.10                               & 0.14                               & 0.17                                & 0.18                             & 0.23                             \\ \cmidrule(l){2-7} 
                                                                               & A2C              & 0.21                               & 0.30                               & 0.43                                & 0.31                             & 0.42                             \\ \cmidrule(l){2-7} 
                                                                               & A2C with Entropy & 0.38                               & 0.46                               & 0.53                                & 0.47                             & 0.49                             \\ \midrule
\multirow{3}{*}{\begin{tabular}[c]{@{}c@{}}Eclipse\\ Platform UI\end{tabular}} & ES               & 0.09                               & 0.15                               & 0.19                                & 0.25                             & 0.31                             \\ \cmidrule(l){2-7} 
                                                                               & A2C              & 0.25                               & 0.38                               & 0.51                                & 0.39                             & 0.51                             \\ \cmidrule(l){2-7} 
                                                                               & A2C with Entropy & 0.45                               & 0.69                               & 0.78                                & 0.54                             & 0.59                             \\ \bottomrule
\end{tabular}

\label{table: A2C performance}
\end{table}
The observed performance of RLocator in achieving higher MAP can be interpreted due to two factors: 1) the way we design our reward function, given that we define a function that aims to encourage higher MAP; 2) the inclusion of entropy, as entropy regularization is assumed to enable the model to achieve higher MAP.

Hence, to provide a better understanding of our model, we measure the performance of three different steps of our model separately. The ES-based shortlisting step, the A2C-based RL model (without entropy), and the A2C with entropy model. Due to resource (time and GPU) limitations, we limit our evaluation to half of the total projects in our dataset, i.e., AspectJ, Birt, and Eclipse Platform UI.  We observe a similar trend in those three projects. Thus, we believe our results will follow a similar trend in the remaining projects.

Table~\ref{table: A2C performance} presents the performance of the three choices (i.e., ES, A2C only, and  A2C with Entropy). 
Table~\ref{table: A2C performance} shows that ES archives the baseline performance, which is 53-61\% lower than the A2C with entropy model in terms of MAP and 47-54\% lower in terms of MRR.
We also find that the MRR and MAP of the models without entropy are lower than those of the A2C with entropy models. Table~\ref{table: A2C performance} shows that in terms of MAP, the performance of A2C with entropy models is higher than A2C models by a range of 27.78-34.04\%. 
In MRR and the top 10, the A2C with entropy model achieves higher performance by a range of 11.86-13.56\% and 18.87 - 36\%, respectively.
Such results indicate that entropy could substantially contribute to the model performance regarding MAP, MRR, and Top K.
Moreover, this shows that the use of entropy encourages the RL agent to explore possible alternate policies~\cite{mniha16}; thus, it has a higher chance of getting a better policy for solving the problem in a given environment than the A2C model.
\section{Related Work}
\label{sec:rel_works}
The work most related to our study falls into studies on bug localization techniques.
% , and 2) studies on RL and ranking problems. 
In the following, we
discuss the related work and reflect on how the work compares with ours.
A plethora of work studied how developers localize bugs~\cite{Bhme2017,Sasso2016,Bettenburg2008,Zimmermann2010}.
For example, Böhme et al.~\cite{Bhme2017} studied how developers debug. 
They found that the most popular technique for localizing a bug is forward reasoning, where developers go through each computational step of a failing test case to identify the location. 
% Multiple studies~\cite{Sasso2016, Bettenburg2008, Zimmermann2010} have been conducted to identify the characteristics of a good bug report. 
Zimmermann et al.~\cite{Zimmermann2010} studied the characteristics of a good bug report and
found that test cases and stack traces are one of the most important criteria that makes a good bug report. While these studies focused on developers' manual localization of bugs, our approach examines the automation of the bug localization process for developers.
% \mei{While these studies look at how developers localize bugs, we look at how to automatically localize bugs for developers.}

Several studies offered test case coverage-based solutions for bug localization~\cite{Vancsics2022, Lou2021, Kim2019}. 
Vancsics et al.~\cite{Vancsics2022} proposed a count-based spectrum instead of a hit-based spectrum in the Spectrum-Based Fault Localization (SBFL) tool.
GRACE~\cite{Lou2021} proposed gated graph neural network-based representation learning to improve the SBFL technique. However, these studies mainly utilize test cases to localize bugs, whereas our approach (RLocator) focuses on the bug report for bug localization.

% \subsection{Non-DL based bug localization}
There were several efforts to assess the impact of query reformulation in improving the performance of existing bug localization tools.~\cite{Rahman2021, Florez2021}.
For example, Rahman et al.~\cite{Rahman2021} found that instead of using the full bug report as a full-text query, a reformulated query with some additional expansion performs better.
BugLocator~\cite{Zhou2012} used a revised Vector Space Model (rVSM) to estimate the textual similarity between bug reports and source code files. 
%These studies used bug reports for bug localization. 
% However, none of those approaches are based on reinforcement learning or deep learning.\par

A few studies incorporated the information of program structures such as the Program Dependence Graph (PDG), Data Flow Graph (DFG)~\cite{Li2019}, and Abstract Syntax Tree (AST) for learning source code representation~\cite{Xiao2018, Xiao2019, Li2019, Liang2019, Han2023}.
For example, CAST~\cite{Liang2019} used AST of the source code to extract the semantic information and then used Word2Vec to project the source code and the bug report in the same embedding space. They used a CNN model that measures the similarity between a bug report and source code.  
The model ranks the file based on the calculated similarity.
Hyloc~\cite{Lam2015} incorporated the techniques of IR-based bug localization with deep learning. 
It concatenated the TF-IDF vector of the source code with repository and file-level metadata. 

Other studies applied several deep learning-based approaches for bug localization~\cite{Zhu2021, Zhang2020, Han2023}.
% the Attention mechanism to improve the performance of bug localization approaches~\cite{Zhu2021, Zhang2020}.
DEMOB~\cite{Zhu2021} used attention on ELMo~\cite{peter2018} embedding, whereas KGBugLocator~\cite{Zhang2020} used attention on graph embedding. 
BL-GAN~\cite{Zhu2022} offered a generative adversarial network (GAN) based solution for bug localization. GAN is often seen as a methodology closely related to reinforcement learning, but it diverges from the typical use of the Markov decision process (MDP), a fundamental aspect of reinforcement learning~\cite{Sutton1999, Sutton2018}. Additionally, BL-GAN has limitations in actively learning bug localization from developers' real-time actions. In contrast, we have incorporated developers' feedback directly into the reward function, allowing RLocator to learn from developers' actions. Xie et al.~\cite{Xie2022} employed GAN to create failing test cases, addressing the data imbalance issue within fault localization methods. White et al.~\cite{White1999} utilized reinforcement learning for fault localization in distributed networks. Nonetheless, their approach involves the reinforcement learning agent understanding how to interact with the network which is different from our approach. In our approach, the agent learns to localize bugs from developers' activity. Rezapour et al.~\cite{Rezapour2023} have provided a thorough exploration of reinforcement learning's application in fault localization within power systems. However, their discussed approaches significantly differ from ours. In the context of power systems, the reinforcement learning agent can directly observe phenomenal components (e.g., current, voltage) related to the environment. Moreover, the agents are allowed to probe the environment by changing current or voltage. In contrast, bug localization entails more abstract phenomenal components in the environment (e.g., interacting code blocks) and agents are not allowed to change any code or execute the code.
%\subsection{DL-based bug localization}
Other studies focused on associating commits with bug reports~\cite{Ni2022, Ciborowska2022}.
% Commit-level bug localization focuses on associating commits with bug reports~\cite{Ni2022, Ciborowska2022}.
For example, FBL-BERT~\cite{Ciborowska2022} used CodeBERT embedding for estimating the similarity between source code files and changesets of a commit. 
Based on the similarity, it ranks the suspicious commit.
FLIM~\cite{Liang2022} also used CodeBERT embedding for estimating similarity. However, FLIM works on function-level bug localization. 
% However,  none of these studies uses RL or MDP in their methodology.

Our approach, RLocator, uses deep reinforcement learning for bug localization, differing from previous similarity-based methods. By formulating the problem as a Markov Decision Process (MDP), we directly optimize evaluation measures. Testing on a dataset of 8,316 projects from six popular Apache projects, our results show significant performance improvement.
% Our work differs from previous studies on bug localization since we propose our approach (RLocator) based on deep reinforcement learning.
% The key insight behind RLocator is directly optimizing the evaluation measures, while the prior work focused on similarity-based approaches.
% We formulate the bug localization problem using a Markov Decision Process (MDP) to optimize
% the evaluation measures directly.
% We experimentally evaluate it based on a benchmark dataset of 8,316 projects from six highly popular Apache projects.
% Our results demonstrate that direct optimization of evaluation
% measures considerably contributes to performance improvement of the bug localization problem.
% RL approach for bl
\section{Threats to Validity}
\label{sec:threats}
RLocator has a number of limitations as well. We identify them and discuss how to overcome the limitations below.\\

\noindent
\textbf{Internal Validity.} One limitation of our approach is we are not able to utilize 9\% of our dataset due to the limitation of text-based search. One may point out that we exclude the bug reports where we do not perform well. But our the XGBoost model in our approach automatically identifies them and we say that we would rather not localize the source code files for these bug reports than localize them incorrectly. 
%We have manually checked the bug reports. 
%Our manual analysis also points out that the length of the source code files is the only similar attribute in those data points. In our defence, bug localization systems are not designed to be used autonomously. We expect a developer to always verify the ranked files to solve the bugs. 
Hence, developers need to rely on their manual analysis only for the 9\%.
%Perhaps this is better than generating low accuracy results for 100\% of the data, which will shorten the time for the end-to-end time required for bug resolution. 
Moreover, as a measure of full transparency, we estimate the lower bound of RLocator performance for the 100\% data and show that the difference is negligible.\\

\noindent
\textbf{External Validity.} The primary concern for the external validity of the RLocator evaluation stems from its limitation to a small number of bugs in six varied, real-world open-source projects, potentially impacting its broad applicability. However, those projects are from different domains and used by prior studies~\cite{Xiao2018, Liang2019, Zhu2021, Zhang2020, Lam2015, Wang2020}. Furthermore, the A2C without entropy model was only evaluated on three projects because of the substantial resources required for training—taking about four days on an Nvidia V100 16GB GPU. The uniform outcomes across these projects indicate that similar results could be expected in the remaining projects. Additionally, due to the absence of a replication package, we replicated BL-GAN based on its description in the original study, which may lead to slight performance deviations. Nevertheless, after experimenting with various hyperparameters, we selected a set that achieves comparable performance to that reported in the original study.\\

\noindent
\textbf{Construct Validity.} Finally, our evaluation measures might be one threat to construct validity. 
The evaluation measures may not completely reflect real-world situations. The threat
is mitigated by the fact that the used evaluation measures are well-known~\cite{Ciborowska2022, Liang2022, Ye2014, Zhou2012}
and best available to measure and compare the
performance of information retrieval-based bug localization tools.
\vspace{20pt}
\section{Conclusion}
\label{sec:conclusion}
 In this paper, we propose RLocator, a reinforcement learning-based (RL) technique to rank the source code files where the bug may reside, given the bug report.
 The key contribution of our study is  the formulation of the bug localization problem using the Markov Decision Process (MDP), which helps us to optimize the evaluation measures directly.
 We evaluate RLocator on 8,316 bug reports and find that RLocator performs better than the state-of-the-art techniques when using MAP as an evaluation measure.
 % Also, it has good most of the time when using MRR. 
 Using 91\% bug reports dataset, RLocator outperforms prior tools in all the project in terms of both MAP and MRR.
 When using 100\% data, RLocator outperforms all prior approaches in four of six projects using MAP and two of the six projects using MRR. 
 RLocator can be used along with other bug localization approaches to improve performance. Our results show that RL is a promising avenue for future exploration when it comes to advancing state-of-the-art techniques for bug localization. \tbl{Future research can explore the application of advanced reinforcement learning algorithms in bug localization. Additionally, researchers can investigate how training on larger datasets impacts the performance of tools in low-similarity contexts.}
 
 %achieves the highest MAP comparison to seven other approaches among bug reports where we can find relevant files. We also estimate the worst possible performance of RLocator and find that the performance drop is not substantial. Moreover, the worst possible MAP of RLocator is better than the other approaches in the three projects. 
 %Furthermore, to put RLocator in a real-world scenario, we develop an XGBoost model to identify the characteristics of bugs that are hard to localize for text-retrieval tools. Our findings point out that the length of source code files is one of the primary factors that makes some bugs hard to localize. In the future, we plan to extend our tool to accommodate any source code files. 
 
\vspace{10pt}
\section{Data Availability}
To foster future research in the field, we make a replication package comprising
our dataset and code are publicly available~\cite{replication}.

 % lower limit 
%  However, we could not utilize the full dataset due to limitations in the text-based search method and fixed size limitation of deep learning models. One future study based on this one can be development of an RL agent that can accept variable actions (variable number of a file in a software project). 

% \appendices
% \section*{Acknowledgment}
\balance
\bibliographystyle{IEEEtran}
\bibliography{Bibliography}

\begin{thebibliography}{10}
\providecommand{\url}[1]{#1}
\csname url@samestyle\endcsname
\providecommand{\newblock}{\relax}
\providecommand{\bibinfo}[2]{#2}
\providecommand{\BIBentrySTDinterwordspacing}{\spaceskip=0pt\relax}
\providecommand{\BIBentryALTinterwordstretchfactor}{4}
\providecommand{\BIBentryALTinterwordspacing}{\spaceskip=\fontdimen2\font plus
\BIBentryALTinterwordstretchfactor\fontdimen3\font minus
  \fontdimen4\font\relax}
\providecommand{\BIBforeignlanguage}[2]{{%
\expandafter\ifx\csname l@#1\endcsname\relax
\typeout{** WARNING: IEEEtran.bst: No hyphenation pattern has been}%
\typeout{** loaded for the language `#1'. Using the pattern for}%
\typeout{** the default language instead.}%
\else
\language=\csname l@#1\endcsname
\fi
#2}}
\providecommand{\BIBdecl}{\relax}
\BIBdecl

\bibitem{LaToza2010}
T.~D. LaToza and B.~A. Myers, ``Developers ask reachability questions,'' in
  \emph{Proceedings of the 32nd {ACM}/{IEEE} International Conference on
  Software Engineering - {ICSE} {\textquotesingle}10}.\hskip 1em plus 0.5em
  minus 0.4em\relax {ACM} Press, 2010.

\bibitem{Anvik2005}
J.~Anvik, L.~Hiew, and G.~C. Murphy, ``Coping with an open bug repository,'' in
  \emph{Proceedings of the 2005 {OOPSLA} workshop on Eclipse technology
  {eXchange} - eclipse {\textquotesingle}05}.\hskip 1em plus 0.5em minus
  0.4em\relax {ACM} Press, 2005.

\bibitem{Zhou2012}
J.~Zhou, H.~Zhang, and D.~Lo, ``Where should the bugs be fixed? more accurate
  information retrieval-based bug localization based on bug reports,'' in
  \emph{2012 34th International Conference on Software Engineering
  ({ICSE})}.\hskip 1em plus 0.5em minus 0.4em\relax {IEEE}, Jun. 2012.

\bibitem{Xiao2018}
Y.~Xiao, J.~Keung, K.~E. Bennin, and Q.~Mi, ``Machine translation-based bug
  localization technique for bridging lexical gap,'' \emph{Information and
  Software Technology}, vol.~99, pp. 58--61, Jul. 2018.

\bibitem{Liang2019}
H.~Liang, L.~Sun, M.~Wang, and Y.~Yang, ``Deep learning with customized
  abstract syntax tree for bug localization,'' \emph{{IEEE} Access}, vol.~7,
  pp. 116\,309--116\,320, 2019.

\bibitem{Zhang2020}
J.~Zhang, R.~Xie, W.~Ye, Y.~Zhang, and S.~Zhang, ``Exploiting code knowledge
  graph for bug localization via bi-directional attention,'' in
  \emph{Proceedings of the 28th International Conference on Program
  Comprehension}.\hskip 1em plus 0.5em minus 0.4em\relax {ACM}, Jul. 2020.

\bibitem{Zhu2022}
Z.~Zhu, H.~Tong, Y.~Wang, and Y.~Li, ``{BL}-{GAN}: Semi-supervised bug
  localization via generative adversarial network,'' \emph{{IEEE} Transactions
  on Knowledge and Data Engineering}, pp. 1--14, 2022.

\bibitem{Ciborowska2022}
A.~Ciborowska and K.~Damevski, ``Fast changeset-based bug localization with
  {BERT},'' in \emph{Proceedings of the 44th International Conference on
  Software Engineering}.\hskip 1em plus 0.5em minus 0.4em\relax {ACM}, May
  2022.

\bibitem{Lam2015}
A.~N. Lam, A.~T. Nguyen, H.~A. Nguyen, and T.~N. Nguyen, ``Combining deep
  learning with information retrieval to localize buggy files for bug reports
  (n),'' in \emph{2015 30th {IEEE}/{ACM} International Conference on Automated
  Software Engineering ({ASE})}.\hskip 1em plus 0.5em minus 0.4em\relax {IEEE},
  Nov. 2015.

\bibitem{Wei2017}
Z.~Wei, J.~Xu, Y.~Lan, J.~Guo, and X.~Cheng, ``Reinforcement learning to rank
  with markov decision process,'' in \emph{Proceedings of the 40th
  International {ACM} {SIGIR} Conference on Research and Development in
  Information Retrieval}.\hskip 1em plus 0.5em minus 0.4em\relax {ACM}, Aug.
  2017.

\bibitem{Alejo2010}
O.~Alejo, J.~M. Fernandez-Luna, J.~F. Huete, and R.~Perez-Vazquez, ``Direct
  optimization of evaluation measures in learning to rank using particle
  swarm,'' in \emph{2010 Workshops on Database and Expert Systems
  Applications}.\hskip 1em plus 0.5em minus 0.4em\relax {IEEE}, Aug. 2010.

\bibitem{Xu2017}
J.~Xu, L.~Xia, Y.~Lan, J.~Guo, and X.~Cheng, ``Directly optimize diversity
  evaluation measures,'' \emph{{ACM} Transactions on Intelligent Systems and
  Technology}, vol.~8, no.~3, pp. 1--26, Jan. 2017.

\bibitem{Yue2007}
Y.~Yue, T.~Finley, F.~Radlinski, and T.~Joachims, ``A support vector method for
  optimizing average precision,'' in \emph{Proceedings of the 30th annual
  international {ACM} {SIGIR} conference on Research and development in
  information retrieval}.\hskip 1em plus 0.5em minus 0.4em\relax {ACM}, Jul.
  2007.

\bibitem{Sutton2018}
R.~S. Sutton and A.~G. Barto, \emph{Reinforcement Learning: An
  Introduction}.\hskip 1em plus 0.5em minus 0.4em\relax Cambridge, MA, USA: A
  Bradford Book, 2018.

\bibitem{Garcia2013}
F.~Garcia and E.~Rachelson, ``Markov decision processes,'' in \emph{Markov
  Decision Processes in Artificial Intelligence}.\hskip 1em plus 0.5em minus
  0.4em\relax John Wiley {\&} Sons, Inc., Mar. 2013, pp. 1--38.

\bibitem{GDI2021}
``Gdi: Rethinking what makes reinforcement learning different from supervised
  learning,'' 2021.

\bibitem{Kober2013}
J.~Kober, J.~A. Bagnell, and J.~Peters, ``Reinforcement learning in robotics: A
  survey,'' \emph{The International Journal of Robotics Research}, vol.~32,
  no.~11, pp. 1238--1274, Aug. 2013.

\bibitem{Liang2022}
H.~Liang, D.~Hang, and X.~Li, ``Modeling function-level interactions for
  file-level bug localization,'' \emph{Empirical Software Engineering},
  vol.~27, no.~7, Oct. 2022.

\bibitem{Maystre2023}
L.~Maystre, D.~Russo, and Y.~Zhao, ``Optimizing audio recommendations for the
  long-term: A reinforcement learning perspective,'' 2023.

\bibitem{Chen2018}
M.~Chen, A.~Beutel, P.~Covington, S.~Jain, F.~Belletti, and E.~Chi, ``Top-k
  off-policy correction for a reinforce recommender system,'' 2018.

\bibitem{Yu2021}
C.~Yu, J.~Liu, S.~Nemati, and G.~Yin, ``Reinforcement learning in healthcare: A
  survey,'' \emph{ACM Computing Surveys}, vol.~55, no.~1, p. 1–36, Nov. 2021.

\bibitem{Winter2023}
E.~Winter, D.~Bowes, S.~Counsell, T.~Hall, S.~Haraldsson, V.~Nowack, and
  J.~Woodward, ``How do developers really feel about bug fixing? directions for
  automatic program repair,'' \emph{IEEE Transactions on Software Engineering},
  vol.~49, no.~04, pp. 1823--1841, apr 2023.

\bibitem{Wang2016}
S.~Wang, T.~Liu, and L.~Tan, ``Automatically learning semantic features for
  defect prediction,'' in \emph{Proceedings of the 38th International
  Conference on Software Engineering}.\hskip 1em plus 0.5em minus 0.4em\relax
  {ACM}, May 2016.

\bibitem{Miryeganeh2021}
N.~Miryeganeh, S.~Hashtroudi, and H.~Hemmati, ``{GloBug}: Using global data in
  fault localization,'' \emph{Journal of Systems and Software}, vol. 177, p.
  110961, Jul. 2021.

\bibitem{Kim2020}
Y.~Kim, M.~Kim, and E.~Lee, ``Feature combination to alleviate hubness problem
  of source code representation for bug localization,'' in \emph{2020 27th
  Asia-Pacific Software Engineering Conference ({APSEC})}.\hskip 1em plus 0.5em
  minus 0.4em\relax {IEEE}, Dec. 2020.

\bibitem{Chen2020}
L.~Chen, Z.~Tang, and G.~H. Yang, ``Balancing reinforcement learning training
  experiences in interactive information retrieval,'' in \emph{Proceedings of
  the 43rd International {ACM} {SIGIR} Conference on Research and Development
  in Information Retrieval}.\hskip 1em plus 0.5em minus 0.4em\relax {ACM}, Jul.
  2020.

\bibitem{devlin2019}
J.~Devlin, M.-W. Chang, K.~Lee, and K.~Toutanova, ``{BERT}: Pre-training of
  deep bidirectional transformers for language understanding,'' in
  \emph{Proceedings of the 2019 Conference of the North {A}merican Chapter of
  the Association for Computational Linguistics: Human Language Technologies,
  Volume 1 (Long and Short Papers)}.\hskip 1em plus 0.5em minus 0.4em\relax
  Minneapolis, Minnesota: Association for Computational Linguistics, Jun. 2019,
  pp. 4171--4186.

\bibitem{feng2020}
Z.~Feng, D.~Guo, D.~Tang, N.~Duan, X.~Feng, M.~Gong, L.~Shou, B.~Qin, T.~Liu,
  D.~Jiang, and M.~Zhou, ``{C}ode{BERT}: A pre-trained model for programming
  and natural languages,'' in \emph{Findings of the Association for
  Computational Linguistics: EMNLP 2020}.\hskip 1em plus 0.5em minus
  0.4em\relax Online: Association for Computational Linguistics, Nov. 2020, pp.
  1536--1547.

\bibitem{Wang2020sim}
J.~Wang and Y.~Dong, ``Measurement of text similarity: A survey,''
  \emph{Information}, vol.~11, no.~9, p. 421, Aug. 2020.

\bibitem{Fujimoto2019}
S.~Fujimoto, D.~Meger, and D.~Precup, ``Off-policy deep reinforcement learning
  without exploration,'' in \emph{Proceedings of the 36th International
  Conference on Machine Learning}, ser. Proceedings of Machine Learning
  Research, K.~Chaudhuri and R.~Salakhutdinov, Eds., vol.~97.\hskip 1em plus
  0.5em minus 0.4em\relax PMLR, 09--15 Jun 2019, pp. 2052--2062.

\bibitem{Nguyen2021}
T.~T. Nguyen and V.~J. Reddi, ``Deep reinforcement learning for cyber
  security,'' \emph{{IEEE} Transactions on Neural Networks and Learning
  Systems}, pp. 1--17, 2021.

\bibitem{Liu2022}
C.~Liu, X.~Xia, D.~Lo, Z.~Liu, A.~E. Hassan, and S.~Li, ``{CodeMatcher}:
  Searching code based on sequential semantics of important query words,''
  \emph{{ACM} Transactions on Software Engineering and Methodology}, vol.~31,
  no.~1, pp. 1--37, Jan. 2022.

\bibitem{Chen2016}
T.~Chen and C.~Guestrin, ``{XGBoost},'' in \emph{Proceedings of the 22nd {ACM}
  {SIGKDD} International Conference on Knowledge Discovery and Data
  Mining}.\hskip 1em plus 0.5em minus 0.4em\relax {ACM}, Aug. 2016.

\bibitem{Fang2021}
F.~Fang, J.~Wu, Y.~Li, X.~Ye, W.~Aljedaani, and M.~W. Mkaouer, ``On the
  classification of bug reports to improve bug localization,'' \emph{Soft
  Computing}, vol.~25, no.~11, pp. 7307--7323, Mar. 2021.

\bibitem{Lv2011}
Y.~Lv and C.~Zhai, ``When documents are very long, {BM}25 fails!'' in
  \emph{Proceedings of the 34th international {ACM} {SIGIR} conference on
  Research and development in Information - {SIGIR}
  {\textquotesingle}11}.\hskip 1em plus 0.5em minus 0.4em\relax {ACM} Press,
  2011.

\bibitem{Lewis1998}
D.~D. Lewis, ``Naive (bayes) at forty: The independence assumption in
  information retrieval,'' in \emph{Machine Learning: {ECML}-98}.\hskip 1em
  plus 0.5em minus 0.4em\relax Springer Berlin Heidelberg, 1998, pp. 4--15.

\bibitem{Schroter2010}
A.~Schroter, A.~Schr\"{o}ter, N.~Bettenburg, and R.~Premraj, ``Do stack traces
  help developers fix bugs?'' in \emph{2010 7th {IEEE} Working Conference on
  Mining Software Repositories ({MSR} 2010)}.\hskip 1em plus 0.5em minus
  0.4em\relax {IEEE}, May 2010.

\bibitem{Lv2011_plus}
Y.~Lv and C.~Zhai, ``Lower-bounding term frequency normalization,'' in
  \emph{Proceedings of the 20th {ACM} international conference on Information
  and knowledge management - {CIKM} {\textquotesingle}11}.\hskip 1em plus 0.5em
  minus 0.4em\relax {ACM} Press, 2011.

\bibitem{replication}
\BIBentryALTinterwordspacing
{Anonymous}, ``\BIBforeignlanguage{en}{Rlocator: Reinforcement learning for bug
  localization},'' 2023. [Online]. Available:
  \url{https://zenodo.org/record/7591879}
\BIBentrySTDinterwordspacing

\bibitem{Bagherzadeh2022}
M.~Bagherzadeh, N.~Kahani, and L.~Briand, ``Reinforcement learning for test
  case prioritization,'' \emph{{IEEE} Transactions on Software Engineering},
  vol.~48, no.~8, pp. 2836--2856, Aug. 2022.

\bibitem{Wan2018}
Y.~Wan, Z.~Zhao, M.~Yang, G.~Xu, H.~Ying, J.~Wu, and P.~S. Yu, ``Improving
  automatic source code summarization via deep reinforcement learning,'' in
  \emph{Proceedings of the 33rd {ACM}/{IEEE} International Conference on
  Automated Software Engineering}.\hskip 1em plus 0.5em minus 0.4em\relax
  {ACM}, Sep. 2018.

\bibitem{He2017}
X.~He, L.~Liao, H.~Zhang, L.~Nie, X.~Hu, and T.-S. Chua, ``Neural collaborative
  filtering,'' in \emph{Proceedings of the 26th International Conference on
  World Wide Web}.\hskip 1em plus 0.5em minus 0.4em\relax International World
  Wide Web Conferences Steering Committee, Apr. 2017.

\bibitem{Zhang2014}
H.~Zhang, Y.~Yang, H.~Luan, S.~Yang, and T.-S. Chua, ``Start from scratch,'' in
  \emph{Proceedings of the 22nd {ACM} international conference on
  Multimedia}.\hskip 1em plus 0.5em minus 0.4em\relax {ACM}, Nov. 2014.

\bibitem{Zhu2020}
R.~Zhu, X.~Tu, and J.~X. Huang, ``Deep learning on information retrieval and
  its applications,'' in \emph{Deep Learning for Data Analytics}.\hskip 1em
  plus 0.5em minus 0.4em\relax Elsevier, 2020, pp. 125--153.

\bibitem{Khattab2020}
O.~Khattab and M.~Zaharia, ``{ColBERT},'' in \emph{Proceedings of the 43rd
  International {ACM} {SIGIR} Conference on Research and Development in
  Information Retrieval}.\hskip 1em plus 0.5em minus 0.4em\relax {ACM}, Jul.
  2020.

\bibitem{Hochreiter1997}
S.~Hochreiter and J.~Schmidhuber, ``Long short-term memory,'' \emph{Neural
  Computation}, vol.~9, no.~8, pp. 1735--1780, Nov. 1997.

\bibitem{Pang2017}
L.~Pang, Y.~Lan, J.~Guo, J.~Xu, J.~Xu, and X.~Cheng, ``{DeepRank},'' in
  \emph{Proceedings of the 2017 {ACM} on Conference on Information and
  Knowledge Management}.\hskip 1em plus 0.5em minus 0.4em\relax {ACM}, Nov.
  2017.

\bibitem{Huo2021}
X.~Huo, F.~Thung, M.~Li, D.~Lo, and S.-T. Shi, ``Deep transfer bug
  localization,'' \emph{{IEEE} Transactions on Software Engineering}, vol.~47,
  no.~7, pp. 1368--1380, Jul. 2021.

\bibitem{Huo2016}
X.~Huo, M.~Li, and Z.-H. Zhou, ``Learning unified features from natural and
  programming languages for locating buggy source code,'' in \emph{IJCAI},
  2016.

\bibitem{Hausknecht2015}
M.~J. Hausknecht and P.~Stone, ``Deep recurrent q-learning for partially
  observable mdps,'' \emph{CoRR}, vol. abs/1507.06527, 2015.

\bibitem{Matthew2015}
------, ``Deep recurrent q-learning for partially observable mdps,''
  \emph{ArXiv}, vol. abs/1507.06527, 2015.

\bibitem{Bello2016}
I.~Bello, H.~Pham, Q.~V. Le, M.~Norouzi, and S.~Bengio, ``Neural combinatorial
  optimization with reinforcement learning,'' 2016.

\bibitem{Haarnoja2018}
T.~Haarnoja, A.~Zhou, P.~Abbeel, and S.~Levine, ``Soft actor-critic: Off-policy
  maximum entropy deep reinforcement learning with a stochastic actor,'' in
  \emph{Proceedings of the 35th International Conference on Machine Learning},
  ser. Proceedings of Machine Learning Research, J.~Dy and A.~Krause, Eds.,
  vol.~80.\hskip 1em plus 0.5em minus 0.4em\relax PMLR, 10--15 Jul 2018, pp.
  1861--1870.

\bibitem{Wang2019}
C.~Wang, C.~Xu, X.~Yao, and D.~Tao, ``Evolutionary generative adversarial
  networks,'' \emph{{IEEE} Transactions on Evolutionary Computation}, vol.~23,
  no.~6, pp. 921--934, Dec. 2019.

\bibitem{Rabinovich_2023}
E.~Rabinovich, M.~Vetzler, S.~Ackerman, and A.~Anaby~Tavor, ``Reliable and
  interpretable drift detection in streams of short texts,'' in
  \emph{Proceedings of the 61st Annual Meeting of the Association for
  Computational Linguistics (Volume 5: Industry Track)}.\hskip 1em plus 0.5em
  minus 0.4em\relax Association for Computational Linguistics, 2023.

\bibitem{Islam_2021}
M.~R. Islam and M.~F. Zibran, ``What changes in where?: an empirical study of
  bug-fixing change patterns,'' \emph{ACM SIGAPP Applied Computing Review},
  vol.~20, no.~4, p. 18–34, January 2021.

\bibitem{Aljedaani_2018}
W.~Aljedaani and Y.~Javed, \emph{Bug Reports Evolution in Open Source
  Systems}.\hskip 1em plus 0.5em minus 0.4em\relax Springer International
  Publishing, 2018, p. 63–73.

\bibitem{Ye2014}
X.~Ye, R.~Bunescu, and C.~Liu, ``Learning to rank relevant files for bug
  reports using domain knowledge,'' in \emph{Proceedings of the 22nd {ACM}
  {SIGSOFT} International Symposium on Foundations of Software Engineering -
  {FSE} 2014}.\hskip 1em plus 0.5em minus 0.4em\relax {ACM} Press, 2014.

\bibitem{Lee2018}
J.~Lee, D.~Kim, T.~F. Bissyand{\'{e}}, W.~Jung, and Y.~L. Traon, ``Bench4bl:
  reproducibility study on the performance of {IR}-based bug localization,'' in
  \emph{Proceedings of the 27th {ACM} {SIGSOFT} International Symposium on
  Software Testing and Analysis}.\hskip 1em plus 0.5em minus 0.4em\relax {ACM},
  Jul. 2018.

\bibitem{Dit2011}
B.~Dit, M.~Revelle, M.~Gethers, and D.~Poshyvanyk, ``Feature location in source
  code: a taxonomy and survey,'' \emph{Journal of Software: Evolution and
  Process}, vol.~25, no.~1, pp. 53--95, Nov. 2011.

\bibitem{Moreno2015}
L.~Moreno, G.~Bavota, S.~Haiduc, M.~D. Penta, R.~Oliveto, B.~Russo, and
  A.~Marcus, ``Query-based configuration of text retrieval solutions for
  software engineering tasks,'' in \emph{Proceedings of the 2015 10th Joint
  Meeting on Foundations of Software Engineering}.\hskip 1em plus 0.5em minus
  0.4em\relax {ACM}, Aug. 2015.

\bibitem{Sisman2013}
B.~Sisman and A.~C. Kak, ``Assisting code search with automatic query
  reformulation for bug localization,'' in \emph{2013 10th Working Conference
  on Mining Software Repositories ({MSR})}.\hskip 1em plus 0.5em minus
  0.4em\relax {IEEE}, May 2013.

\bibitem{Xiao2019}
Y.~Xiao, J.~Keung, K.~E. Bennin, and Q.~Mi, ``Improving bug localization with
  word embedding and enhanced convolutional neural networks,''
  \emph{Information and Software Technology}, vol. 105, pp. 17--29, Jan. 2019.

\bibitem{Schwarz1981}
M.~N. Schwarz and A.~Flammer, ``Text structure and title{\textemdash}effects on
  comprehension and recall,'' \emph{Journal of Verbal Learning and Verbal
  Behavior}, vol.~20, no.~1, pp. 61--66, Feb. 1981.

\bibitem{Han2023}
J.~Han, C.~Huang, S.~Sun, Z.~Liu, and J.~Liu, ``{bjXnet}: an improved bug
  localization model based on code property graph and attention mechanism,''
  \emph{Automated Software Engineering}, vol.~30, no.~1, Mar. 2023.

\bibitem{Cheng2020}
S.~Cheng, X.~Yan, and A.~A. Khan, ``A similarity integration method based
  information retrieval and word embedding in bug localization,'' in \emph{2020
  {IEEE} 20th International Conference on Software Quality, Reliability and
  Security ({QRS})}.\hskip 1em plus 0.5em minus 0.4em\relax {IEEE}, Dec. 2020.

\bibitem{Soltani2020}
M.~Soltani, F.~Hermans, and T.~B\"{a}ck, ``The significance of bug report
  elements,'' \emph{Empirical Software Engineering}, vol.~25, no.~6, p.
  5255–5294, Sep. 2020.

\bibitem{Ahmed2019}
Z.~Ahmed, N.~Le~Roux, M.~Norouzi, and D.~Schuurmans, ``Understanding the impact
  of entropy on policy optimization,'' in \emph{Proceedings of the 36th
  International Conference on Machine Learning}, ser. Proceedings of Machine
  Learning Research, K.~Chaudhuri and R.~Salakhutdinov, Eds., vol.~97.\hskip
  1em plus 0.5em minus 0.4em\relax PMLR, 09--15 Jun 2019, pp. 151--160.

\bibitem{Jang2022}
S.~Jang and H.-I. Kim, ``Entropy-aware model initialization for effective
  exploration in deep reinforcement learning,'' \emph{Sensors}, vol.~22,
  no.~15, p. 5845, Aug. 2022.

\bibitem{mniha16}
V.~Mnih, A.~P. Badia, M.~Mirza, A.~Graves, T.~Lillicrap, T.~Harley, D.~Silver,
  and K.~Kavukcuoglu, ``Asynchronous methods for deep reinforcement learning,''
  in \emph{Proceedings of The 33rd International Conference on Machine
  Learning}, ser. Proceedings of Machine Learning Research, M.~F. Balcan and
  K.~Q. Weinberger, Eds., vol.~48.\hskip 1em plus 0.5em minus 0.4em\relax New
  York, New York, USA: PMLR, 20--22 Jun 2016, pp. 1928--1937.

\bibitem{Bhme2017}
M.~B\"{o}hme, E.~O. Soremekun, S.~Chattopadhyay, E.~Ugherughe, and A.~Zeller,
  ``Where is the bug and how is it fixed? an experiment with practitioners,''
  in \emph{Proceedings of the 2017 11th Joint Meeting on Foundations of
  Software Engineering}.\hskip 1em plus 0.5em minus 0.4em\relax {ACM}, Aug.
  2017.

\bibitem{Sasso2016}
T.~D. Sasso, A.~Mocci, and M.~Lanza, ``What makes a satisficing bug report?''
  in \emph{2016 {IEEE} International Conference on Software Quality,
  Reliability and Security ({QRS})}.\hskip 1em plus 0.5em minus 0.4em\relax
  {IEEE}, Aug. 2016.

\bibitem{Bettenburg2008}
N.~Bettenburg, S.~Just, A.~Schr\"{o}ter, C.~Weiss, R.~Premraj, and
  T.~Zimmermann, ``What makes a good bug report?'' in \emph{Proceedings of the
  16th {ACM} {SIGSOFT} International Symposium on Foundations of software
  engineering}.\hskip 1em plus 0.5em minus 0.4em\relax {ACM}, Nov. 2008.

\bibitem{Zimmermann2010}
T.~Zimmermann, R.~Premraj, N.~Bettenburg, S.~Just, A.~Schroter, and C.~Weiss,
  ``What makes a good bug report?'' \emph{{IEEE} Transactions on Software
  Engineering}, vol.~36, no.~5, pp. 618--643, Sep. 2010.

\bibitem{Vancsics2022}
B.~Vancsics, F.~Horv{\'{a}}th, A.~Szatm{\'{a}}ri, and {\'{A}}.~Besz{\'{e}}des,
  ``Fault localization using function call frequencies,'' \emph{Journal of
  Systems and Software}, vol. 193, p. 111429, Nov. 2022.

\bibitem{Lou2021}
Y.~Lou, Q.~Zhu, J.~Dong, X.~Li, Z.~Sun, D.~Hao, L.~Zhang, and L.~Zhang,
  ``Boosting coverage-based fault localization via graph-based representation
  learning,'' in \emph{Proceedings of the 29th {ACM} Joint Meeting on European
  Software Engineering Conference and Symposium on the Foundations of Software
  Engineering}.\hskip 1em plus 0.5em minus 0.4em\relax {ACM}, Aug. 2021.

\bibitem{Kim2019}
Y.~Kim, S.~Mun, S.~Yoo, and M.~Kim, ``Precise learn-to-rank fault localization
  using dynamic and static features of target programs,'' \emph{{ACM}
  Transactions on Software Engineering and Methodology}, vol.~28, no.~4, pp.
  1--34, Oct. 2019.

\bibitem{Rahman2021}
M.~M. Rahman, F.~Khomh, S.~Yeasmin, and C.~K. Roy, ``The forgotten role of
  search queries in {IR}-based bug localization: an empirical study,''
  \emph{Empirical Software Engineering}, vol.~26, no.~6, Aug. 2021.

\bibitem{Florez2021}
J.~M. Florez, O.~Chaparro, C.~Treude, and A.~Marcus, ``Combining query
  reduction and expansion for text-retrieval-based bug localization,'' in
  \emph{2021 {IEEE} International Conference on Software Analysis, Evolution
  and Reengineering ({SANER})}.\hskip 1em plus 0.5em minus 0.4em\relax {IEEE},
  Mar. 2021.

\bibitem{Li2019}
Y.~Li, S.~Wang, T.~N. Nguyen, and S.~V. Nguyen, ``Improving bug detection via
  context-based code representation learning and attention-based neural
  networks,'' \emph{Proceedings of the {ACM} on Programming Languages}, vol.~3,
  no. {OOPSLA}, pp. 1--30, Oct. 2019.

\bibitem{Zhu2021}
Z.~Zhu, Y.~Li, Y.~Wang, Y.~Wang, and H.~Tong, ``A deep multimodal model for bug
  localization,'' \emph{Data Mining and Knowledge Discovery}, Apr. 2021.

\bibitem{peter2018}
M.~E. Peters, M.~Neumann, M.~Iyyer, M.~Gardner, C.~Clark, K.~Lee, and
  L.~Zettlemoyer, ``Deep contextualized word representations,'' \emph{CoRR},
  vol. abs/1802.05365, 2018.

\bibitem{Sutton1999}
R.~S. Sutton, D.~Precup, and S.~Singh, ``Between {MDPs} and semi-{MDPs}: A
  framework for temporal abstraction in reinforcement learning,''
  \emph{Artificial Intelligence}, vol. 112, no. 1-2, pp. 181--211, Aug. 1999.

\bibitem{Xie2022}
H.~Xie, Y.~Lei, M.~Yan, Y.~Yu, X.~Xia, and X.~Mao, ``A universal data
  augmentation approach for fault localization,'' in \emph{Proceedings of the
  44th International Conference on Software Engineering}.\hskip 1em plus 0.5em
  minus 0.4em\relax {ACM}, May 2022.

\bibitem{White1999}
T.~White and B.~Pagurek, ``Distributed fault location in networks using
  learning mobile agents,'' in \emph{Approaches to Intelligence Agents}.\hskip
  1em plus 0.5em minus 0.4em\relax Springer Berlin Heidelberg, 1999, pp.
  182--196.

\bibitem{Rezapour2023}
H.~Rezapour, S.~Jamali, and A.~Bahmanyar, ``Review on artificial
  intelligence-based fault location methods in power distribution networks,''
  \emph{Energies}, vol.~16, no.~12, p. 4636, Jun. 2023.

\bibitem{Ni2022}
C.~Ni, W.~Wang, K.~Yang, X.~Xia, K.~Liu, and D.~Lo, ``The best of both worlds:
  integrating semantic features with expert features for defect prediction and
  localization,'' in \emph{Proceedings of the 30th {ACM} Joint European
  Software Engineering Conference and Symposium on the Foundations of Software
  Engineering}.\hskip 1em plus 0.5em minus 0.4em\relax {ACM}, Nov. 2022.

\bibitem{Wang2020}
B.~Wang, L.~Xu, M.~Yan, C.~Liu, and L.~Liu, ``Multi-dimension convolutional
  neural network for bug localization,'' \emph{{IEEE} Transactions on Services
  Computing}, pp. 1--1, 2020.

\end{thebibliography}
\balance
\begin{IEEEbiography}[{\includegraphics[width=1in,height=0.7in,clip,keepaspectratio]{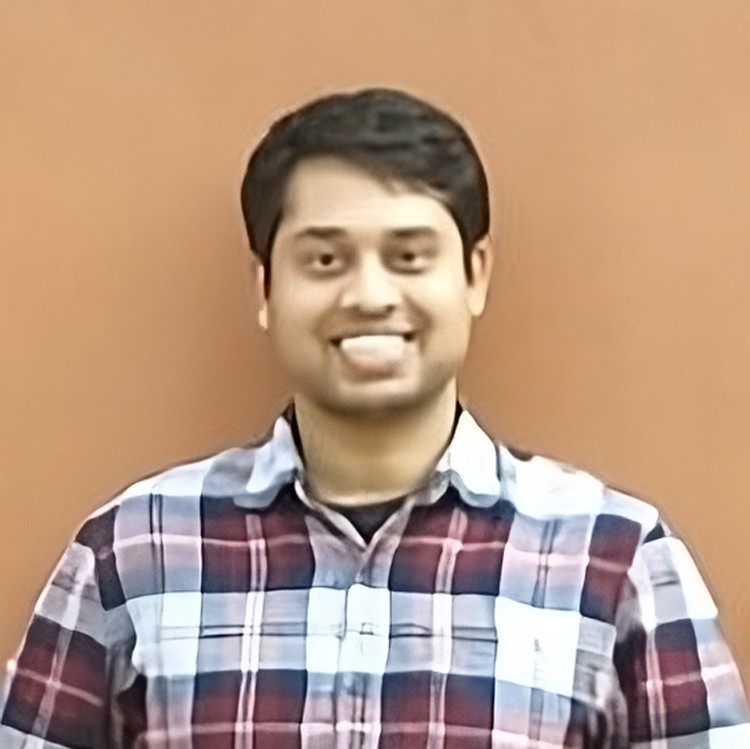}}]{Partha Chakraborty} is a Ph.D. candidate in the
Cheriton School of Computer Science at the
University of Waterloo, Canada. His research
interests include bug localization, vulnerability detection, and the use of machine learning techniques in software engineering. Find more about him at \url{https://parthac.me/.}
\end{IEEEbiography}

\begin{IEEEbiography}[{\includegraphics[width=1in,height=0.7in,clip,keepaspectratio]{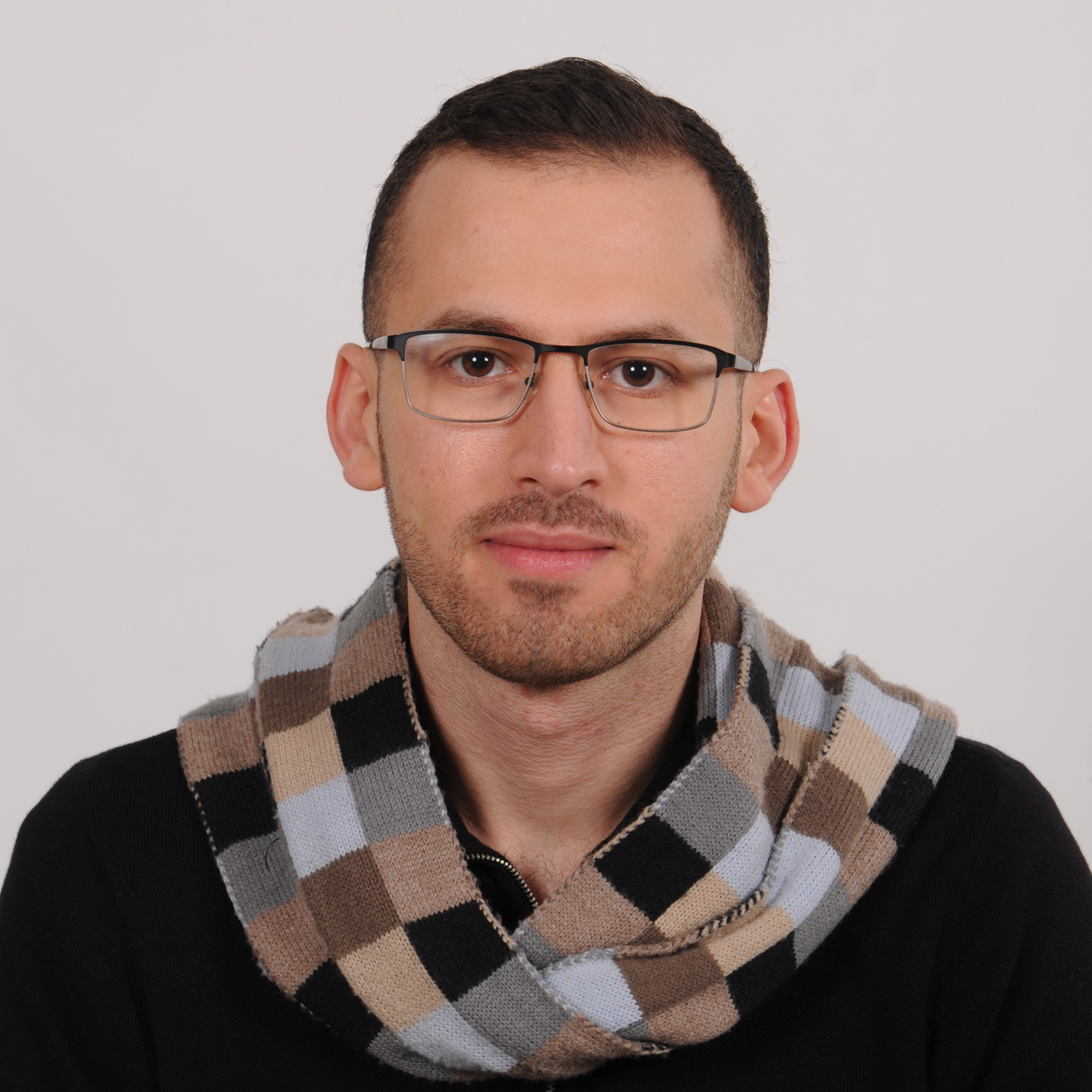}}]{Mahmoud Alfadel}  is an Assistant Professor at the Department of Computer Science, University of Calgary. His research interests include mining software repositories, software ecosystems, open-source security, and release engineering.
\end{IEEEbiography}

\begin{IEEEbiography}[{\includegraphics[width=1in,height=0.7in,clip,keepaspectratio]{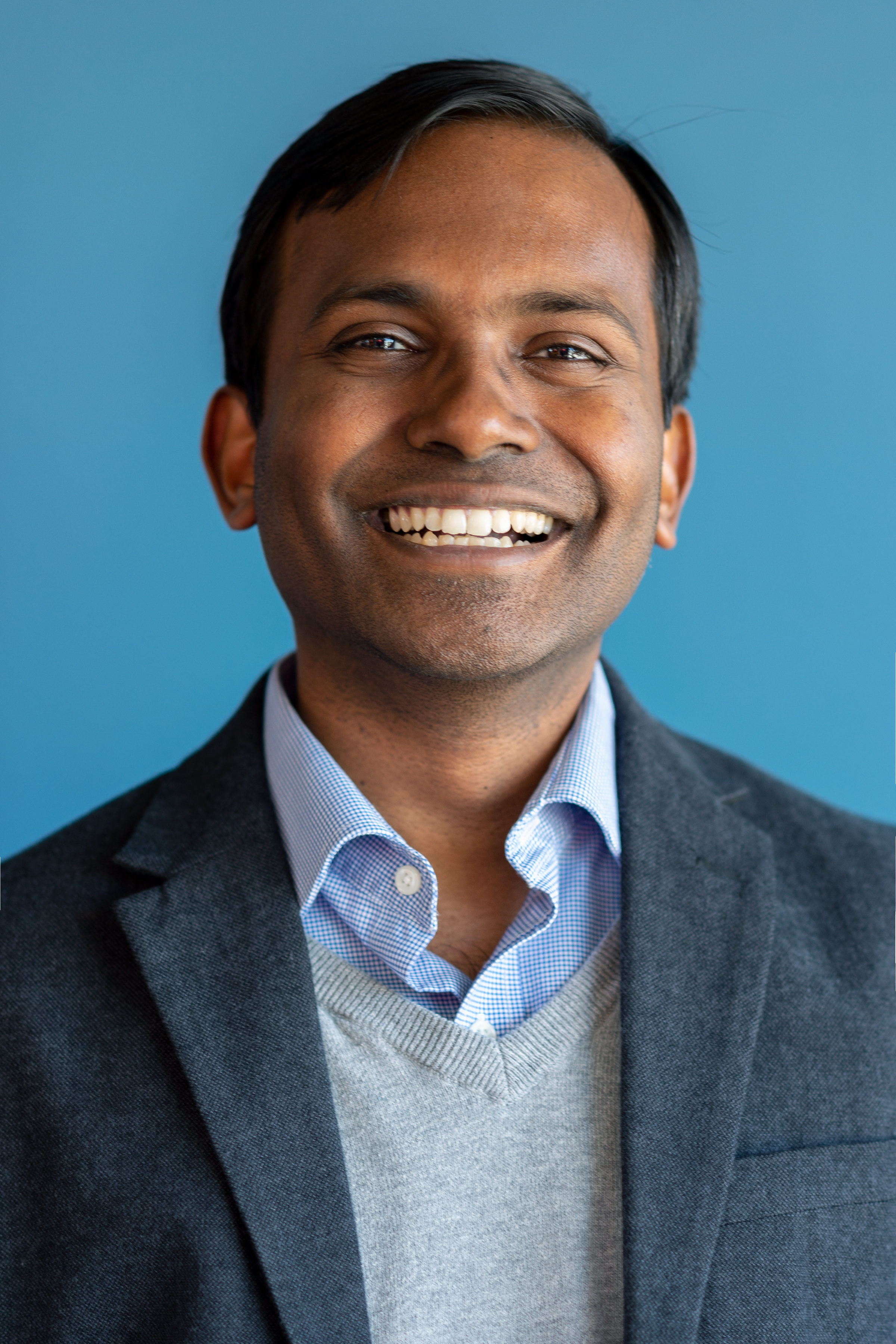}}]{Meiyappan Nagappan} is an Associate Professor at the Cheriton School of Computer Science, University of Waterloo. He has worked on empirical software engineering to address software development concerns and currently researches the impact of large language models on software development.
\end{IEEEbiography}
\vskip -5\baselineskip plus -1fil
\end{document}